\documentclass[useAMS,usenatbib]{mn2e}
\usepackage{amssymb}
\usepackage{amsmath} 
\usepackage{aas_macros}
\usepackage{natbib}
\usepackage{times}
\usepackage{graphicx}
\usepackage{color}

\newcommand{\Mpc}                      {\,{\rm Mpc}}
\newcommand{\Msun}                    {\,{\rm M}_\odot}

\newcommand{\hkpc}                     {\,h^{-1}\,{\rm kpc}}
\newcommand{\hMpc}                    {\,h^{-1}\,{\rm Mpc}}
\newcommand{\hMsun}                  {\,h^{-1}\,{\rm M}_\odot}

\newcommand{\newl} 	{\mathrm{log_{10}}}

\def\lsim{\mathrel{\lower0.6ex\hbox{$\buildrel {\textstyle <}
 \over {\scriptstyle \sim}$}}}

\title[Intrinsic Alignments in EAGLE and COSMO-OWLS]{Intrinsic alignments of galaxies in the EAGLE and cosmo-OWLS simulations}%
 \author[M. Velliscig et al.]{Marco Velliscig,$^{1}$\thanks{E-mail: velliscig@strw.leidenuniv.nl} 
Marcello Cacciato,$^{1}$   
Joop Schaye,$^{1}$ 
Henk Hoekstra,$^{1}$\newauthor
Richard G. Bower,$^2$  
Robert A. Crain,$^{3}$ 
Marcel P. van Daalen,$^{1,4,5}$ \newauthor
Michelle Furlong,$^2$ 
I.~G. McCarthy,$^3$ 
Matthieu Schaller,$^2$ 
Tom Theuns$^{2}$ 
\\\\
$^1$Leiden Observatory, Leiden University, P.O. Box 9513, 2300 RA Leiden, The Netherlands \\
$^2$ Institute for Computational Cosmology, Department of Physics, University of Durham, South Road, Durham, DH1 3LE, UK\\
$^3$Astrophysics Research Institute, Liverpool John Moores University, 146 Brownlow Hill, Liverpool L3 5RF \\
$^4$Max Planck Institute for Astrophysics, Karl-Schwarzschild Stra\ss{}e 1, 85741 Garching, Germany\\
$^5$Department of Astronomy, Theoretical Astrophysics Center, and Lawrence Berkeley National Laboratory, University of California, Berkeley, CA 94720, USA \\
}

\begin{document}

\date{\today}
\pagerange{\pageref{firstpage}--\pageref{lastpage}} \pubyear{2015}

\maketitle

\label{firstpage}
\begin{abstract}
We report results for the alignments of galaxies in the EAGLE and cosmo-OWLS hydrodynamical cosmological simulations as a function of galaxy separation ($-1 \le \newl(r/[\hMpc]) \le 2$) and halo mass ($10.7 \le \newl({M_{\rm 200}/ [\hMsun]}) \le 15$).
We focus on two classes of alignments: the orientations of galaxies with respect to either the {\it directions} to, or the {\it orientations} of, surrounding galaxies.
We find that the strength of the alignment is a strongly decreasing function of the distance between galaxies.
For galaxies hosted by the most massive haloes in our simulations the alignment can remain significant up to $\sim100 \Mpc$.
Galaxies hosted by more massive haloes show stronger alignment.
At a fixed halo mass, more aspherical or prolate galaxies exhibit stronger alignments.
The spatial distribution  of  satellites  is  anisotropic  and  significantly aligned with the major axis of the main host halo.
The major axes of satellite galaxies, when all stars are considered, are preferentially aligned towards the centre of the main host halo.
The predicted projected direction-orientation alignment, $\epsilon_{\rm g+}(r_{\rm p})$, is in broad agreement with recent observations. 
We find that the orientation-orientation alignment is weaker than the orientation-direction alignment on all scales.
Overall, the strength of galaxy alignments depends strongly on the subset of stars that are used to measure the orientations of galaxies and it is always weaker than the alignment of dark matter haloes. Thus, alignment models that use halo orientation as a direct proxy for galaxy orientation  overestimate the impact of intrinsic galaxy alignments.
\end{abstract}


\section{Introduction}
\label{Sec:introduction}

Tidal gravitational fields generated by the formation and evolution of large-scale structures tend to align galaxies due to correlations of tidal torques in random gaussian fields \citep[e.g.][]{Heavens88}.
Analytic theories have been developed to describe these large-scale alignments \citep[linear alignment theory;][]{Catelan00}, but these are only applicable to low matter density contrasts (the linear regime of structure formation) and do not account for drastic events such as mergers of structures, which may erase initial correlations.

To overcome these limitations, galaxy alignments have been studied via N-body simulations \citep[see e.g.][]{West91, Tormen97, Croft00, Heavens00, Jing02, Lee08, Bett12}.
The  most common ansatz in such studies is that galaxies are perfectly aligned with their dark matter haloes and that one can therefore translate the alignments of haloes directly into those of the galaxies that they host. 
However, the observed light from galaxies is emitted by the baryonic component of haloes and hydro-dynamical simulations of galaxy formation have revealed a misalignment between the baryonic and dark matter components of haloes \citep[][]{Deason11,Tenneti14a,Velliscig15}. 
On spatial scales characteristic of a galaxy, baryon processes (radiative cooling, supernova explosions and AGN feedback) play an important role in shaping the spatial distribution of the stars that constitute a galaxy. 
Specifically, the ratio between cooling and heating determines the way baryons lose angular momentum and consequently the way they settle inside their dark matter haloes. 
Furthermore, feedback from star formation and AGN can heat and displace large quantities of gas and inhibit star formation \citep[][]{Springel05_agn, DiMatteo05, DiMatteo08, Booth09, McCarthy10}. 
These processes, which determine when and where stars form, may influence the observed morphology of galaxies and in turn their observed orientations. 
In hydrodynamical simulations of galaxy formation in a cosmological volume, such processes are modelled simultaneously, leading to a potentially more realistic realization of galaxy alignments. The study of such models can unveil patterns that encode important information concerning both the initial conditions that gave rise to the large-scale structure, and the evolution of highly non-linear structures like groups and clusters of galaxies. 

Beyond their relevance to galaxy formation theory, galaxy alignments are a potential contaminant of weak gravitational lensing measurements. Although this contamination is relatively mild, it is a significant concern for large-area cosmic shear surveys \citep[][and references therein]{Joachimi15,Kirk15,Kiessling15}.
Requirements for the precision and accuracy of such surveys are very challenging, as their main goal is to constrain the dark energy equation-of-state parameters at the sub-percent level. 
Weak lensing surveys are used to measure the effect of the bending of light paths of photons emitted from distant galaxies due to intervening matter density contrasts along the line of sight. 
The distortion and magnification of galaxy images is so weak that it can only be characterized by correlating the shapes and orientations of  large numbers of background galaxies. 
In a pure weak gravitational lensing setting, the observed ellipticity of a galaxy, $\epsilon$, is the sum of the intrinsic shape of the galaxy, $\epsilon^s$, and the shear distortion that the light of the galaxy experiences due to gravitational lensing, $\gamma$, 
\begin{equation}
\epsilon = \epsilon^s + \gamma \, .
\label{eq:e_s}
\end{equation}
If galaxies are randomly oriented, the average ellipticity of a sample of galaxies, $\left<\epsilon^s\right>$, vanishes. 
Therefore, any detection of a nonzero $\left<\epsilon\right>$ is interpreted as a measurement of gravitational shear $ \gamma$. However, in the limit of a very weak lensing signal, the distortion induced via gravitational forces (giving rise to an {\it intrinsic} alignment) can be a non-negligible fraction of the distortion due to the pure gravitational lensing effect (often termed {\it apparent} alignment, see \citealt{Crittenden01} and \citealt{Crittenden02} for a statistical description of this effect).

Cosmic shear measurements are obtained in the form of projected 2-point correlation functions (or their equivalent angular power spectra) between shapes of galaxies. Following Eq.~\ref{eq:e_s}:
\begin{align}
\label{eq:eps_eps_corr}
\left<\epsilon\epsilon\right> &= \left<\gamma \gamma\right> + \left<\gamma \epsilon^s\right>+ \left<\epsilon^s \gamma\right>  +\left<\epsilon^s\epsilon^s\right> \, ,\\
&= GG + GI + IG + II \, .
\end{align}
If we assume that galaxies are not intrinsically oriented towards one another, then the only correlations in the shape and orientation of observed galaxies is due to the gravitational lensing effect of the intervening mass distribution between the sources and the observer, $\left<\gamma \gamma\right>$. In this case the only nonzero term is the GG (shear-shear) auto correlation.
In the case of a non negligible intrinsic alignment of galaxies, the II term is also nonzero, i.e. part of the correlation between the shape and orientation of galaxies is \emph{intrinsic}.
If the same gravitational forces that shear the light emitted from a galaxy also tidally influence the intrinsic shape of other galaxies, then this will produce a nonzero cross correlation between shear and intrinsic shape (GI). 
The  term IG is zero since a foreground galaxy cannot be lensed by the same structure that is tidally influencing a background galaxy, unless their respective position along the line of sight is confused due to large errors in the redshift measurements. 

In this paper we report results for the {\it intrinsic} alignment of galaxies in hydro-cosmological simulations. Specifically, we focus on the orientation-direction and  orientation-orientation galaxy alignments. To this aim, we define as galaxy orientation the major eigenvector of the inertia tensor of the distribution of stars in the subhalo. We then compute the mean values of the angle  between the galaxy orientation and the separation vector of other galaxies, as a function of their distance. In the case of orientation-orientation alignment we compute the mean value of the angle between the major axes of the galaxy pairs, as a function of their distance.
While the orientation-orientation alignment can be interpreted straightforwardly as the II term in Eq.~(\ref{eq:eps_eps_corr}), the orientation-direction is related to the GI term in a less direct way \citep[see][for a derivation of the GI power spectrum from the ellipticity correlation function]{Joachimi11}. 

In this paper we make use of four complementary simulations to explore the dependence of the orientation-direction alignment over four orders of magnitude in subhalo mass, and spanning physical separations of hundreds of Mpc. The use of four simulations of different cosmological volumes offer both resolution and statistics, whilst also incorporating baryon physics. The EAGLE simulations used in this work have been calibrated to reproduce the observed present-day galaxy stellar mass function and the observed size-mass relation of disc galaxies (\citealt{Schaye14}), whereas the cosmo-OWLS \citep{LeBrun14, McCarthy14} simulations reproduce key (X-ray and optical) observed properties of galaxy groups and clusters, in addition to the observed galaxy mass function for haloes more massive than $\log(M/[\hMsun]) = 13$. In \citet{Velliscig15} we used the same set of simulations to study the shape and relative alignment of the distributions of stars, dark matter, and hot gas within their own host haloes.  One of the conclusions was that although galaxies align relatively well with the local distribution of the total (mostly dark) matter, they exhibit much larger misalignments with respect to the orientiation of their complete host haloes.

After the submission of this manuscript, a paper by \citet{Chisari15} appeared on the arXiv. They study the alignment of galaxies at $z=0.5$ in the  cosmological hydrodynamical simulation \textsc{Horizon}-AGN \citep[][]{Dubois14} run with the adaptive-mesh-refinement code \textsc{RAMSES} \citep[][]{Teyssier02}. The \textsc{Horizon}-AGN simulation is run in a $(100\hMpc)^3$ volume with a dark matter particle mass resolution of $m_{\rm dm} = 8 \times 10^7 \Msun$. They focus on a galaxy stellar mass range of $9<\newl(M_{\rm star}/[\Msun])<12.36$ and separations up to $25\hMpc$. Their analysis differs in various technical, as well as conceptual, aspects from the study presented here. However, they also report that the strength of galaxy alignments depends strongly on the subset of stars that are used to measure the orientations of galaxies, as found in our investigation.

\begin{table*}
\begin{minipage}{168mm}
\begin{center}
\caption{List of the simulations used and their relevant properties. Description of the columns: (1) descriptive simulation name; (2) comoving size of the simulation box; (3) total number of particles; (4) cosmological parameters; (5) initial mass of baryonic particles; (6) mass of dark matter particles; (7) maximum proper softening length; (8) simulation name tag.} 
\begin{tabular}{llrccccl}
\hline
Simulation  & L & $N_{\rm particle}$& Cosmology & $m_{\mathrm{b}}  $ & $m_{\mathrm{dm}}$ & $\epsilon_{\rm prop}$ & tag \\
 & & &  & $ [\hMsun] $ & $[\hMsun]$ & $[\hkpc]$ &  \\
 (1)&(2)& (3)& (4)& (5)& (6)& (7)&(8) \\ 
\hline 
{EAGLE Recal} &  \,\,\,25 $[\rm Mpc]$   & $2 \times 752^3$  & PLANCK & $ 1.5 \times 10^5$  & $ 8.2 \times 10^5$ & $0.2$     & EA L025\\
{EAGLE Ref}    & 100 $[\rm Mpc]$   & $2 \times 1504^3$ & PLANCK & $ 1.2 \times 10^6$  & $ 6.6  \times 10^6$ & $0.5$   & EA L100\\
{cosmo-OWLS AGN 8.0}      & 200 $[\hMpc]$ & $2 \times 1024^3$ & WMAP7  & $ 8.7  \times 10^7$  & $ 4.1  \times 10^8$ & $2.0$  & CO L200\\
{cosmo-OWLS AGN 8.0}      & 400 $[\hMpc]$ & $2 \times 1024^3$ & WMAP7  & $ 7.5  \times 10^8$  & $ 3.7  \times 10^9$ & $4.0$  & CO L400\\
\hline
\end{tabular}
\label{tbl:simsIA} 
\end{center}
\end{minipage}
\end{table*}

Throughout the paper, we assume a flat $\Lambda$CDM cosmology with massless neutrinos. 
Such a cosmological model is characterized by five parameters: \{$\Omega_{ \rm m}$, $\Omega_{ \rm b}$, $\sigma_{ \rm 8}$, $n_{ \rm s}$, $h$\}. 
The EAGLE and cosmo-OWLS simulations were run with two slightly different sets of values for these parameters.
Specifically, EAGLE was run using the set of cosmological values suggested by the Planck mission \{$\Omega_{ \rm m}$,\, $\Omega_{ \rm b}$,\,$\sigma_{ \rm 8}$,\, $n_{ \rm s}$,\, $h$\} = \{0.307, 0.04825, 0.8288, 0.9611, 0.6777\} (Table 9; \citealt{Planck13}), whereas cosmo-OWLS was run using the cosmological parameters suggested by the 7th-year data release \citep{wmap7} of the WMAP mission \{$\Omega_{ \rm m}$, $\Omega_{ \rm b}$, $\sigma_{ \rm 8}$, $n_{ \rm s}$, $h$\} = \{0.272, 0.0455, 0.728, 0.81, 0.967, 0.704\}.

This paper is organized as follows. In Section~\ref{Sec:methods} we summarize the properties of the simulations employed in this study (\S~\ref{Sec:simulations}) and we introduce the technical definitions used throughout the paper (\S~\ref{Sec:methods_halodef} and \S~\ref{Sec:ShapeParameters}). 
In Section~\ref{Sec:Result_orientation_direction} we report the dependence of the orientation-direction alignment of galaxies on subhalo mass  (\S~\ref{Sec:auto_cross_mass_cos_phi}), matter components (\S~\ref{Sec:total_dm_star_halfstar_cos_phi}), galaxy morphology (\S~\ref{Sec:morpho_cos_phi}) and subhalo type (\S~\ref{Sec:CenSat}). 
In Section~\ref{Sec:obs} we compare our results with observations of the orientation-direction alignment. 
In section \ref{Sec:II} we report results for the orientation-orientation alignment of galaxies.
We summarize our findings and conclude in Section~\ref{Sec:Conclusions}.

\section{Simulations and Technical Definitions}
\label{Sec:methods}

\subsection{Simulations}\label{Sec:simulations}
In this work we employ two different sets of hydrodynamical cosmological simulations, EAGLE \citep{Schaye14, Crain15} and cosmo-OWLS \citep{Schaye10, LeBrun14, McCarthy14}. Specifically, from the EAGLE project we make use of the simulations run in domains of boxsize $L=25$ and $100$ comoving $\Mpc$ in order to study with sufficient resolution central and satellite galaxies hosted by subhaloes with mass from $\newl(M_{\rm sub}/[\hMsun])=10.7$ up to $\newl(M_{\rm sub}/[\hMsun])=12.6$, whereas from cosmo-OWLS we select the simulations run in domains of boxsize $L=200$ and $400$ comoving $\hMpc$ which enable us to extend our analysis to $\newl(M_{\rm sub}/[\hMsun])=15$. 
For each simulation the minimum value of subhalo mass is chosen to be the subhalo mass above which all haloes have at least 300 stellar particles. Using 300 particles ensures a reliable estimation of the subhalo shape \citep[][]{Velliscig15}.
Table~\ref{tbl:simsIA} lists relevant specifics of these simulations. A relevant feature of our composite sample of haloes, taken from four different simulations, is that it reproduces the stellar mass halo mass relation inferred from abundance matching techniques studies \citep{Schaye14}, which ensures that galaxies in our simulations reside in subhaloes of the right mass.

EAGLE and cosmo-OWLS were both run using modified versions of the $N$-Body Tree-PM smoothed particle hydrodynamics (SPH) code \textsc{gadget}~3~\citep{Springel05_gadget}. The simulations employed in this work make use of element-by-element radiative cooling \citep{Wiersma09a}, star formation \citep[][]{Schaye08}, stellar mass losses \citep{Wiersma09b}, stellar feedback \citep{DallaVecchia08,DallaVecchia12}, Black Hole (BH) growth through gas accretion and mergers \citep{Booth09, Rosas13}, and thermal AGN feedback \citep{Booth09, Schaye14}. 

The subgrid physics used in cosmo-OWLS is identical to that used in the OWLS run "AGN" \citep[][]{Schaye10}.
EAGLE includes a series of developments with respect to cosmo-OWLS in the subgrid physics, namely the use of thermal \citep[][]{DallaVecchia12}, instead of kinetic, energy feedback from star formation, BH accretion that depends on the gas angular momentum \citep[][]{Rosas13} and a metallicity dependent star formation law. 
More information regarding the technical implementation of EAGLE's hydro-dynamical aspects, as well as the subgrid physics, can be found in \citet{Schaye14}.

\subsection{Halo and subhalo definition}
\label{Sec:methods_halodef}
Haloes are identified by first applying the Friends-of-Friends (FoF) algorithm to the dark matter particles, with linking length $0.2$ \citep{Davis85}. Baryonic particles are associated to their closest dark matter particle and they inherit their group classification.
Subhaloes are identified as groups of particles in local minima of the gravitational potential. The gravitational potential is calculated for the different particle types separately and then added in order to avoid biases due to different particle masses. Local minima are identified by locating saddle points in the gravitational potential. All particles bound to a given local minimum constitute a subhalo.
The most massive subhalo in a given halo is the \emph{central} subhalo, whereas the others are \emph{satellite} subhaloes. Minima of the gravitational potential are used to identify the centers of subhaloes.
The subhalo mass $M_{\rm sub}$ is the sum of the masses of all the particles belonging to the subhalo.
For every subhalo we define the radius $r_{\rm half}^{\rm dm}$ within which half the mass in dark matter is found.
Similarly, but using stellar particles, we define $r_{\rm half}^{\rm star}$ (usually around one order of magnitude smaller than $r_{\rm half}^{\rm dm}$), which represents a proxy for the typical observable extent of a galaxy within a subhalo.
The $r_{200}^{\rm crit}$ is the radius of the sphere, centered on the central subhalo, that encompasses a mean density that is 200 times the critical density of the Universe. The mass within $r_{200}^{\rm crit}$ is the halo mass $M_{200}^{\rm crit}$.
The aforementioned quantities are computed using \textsc{subfind} \citep{Springel01_subfind,Dolag09}. 

In Table~\ref{tbl:sims_stat_IA} we summarize the $z=0$ values of various quantities of interest for the halo mass bins analysed here.

\begin{table*} 
\begin{minipage}{168mm}
\begin{center}
\caption{Values at $z=0$ of various quantities of interest for our four subhalo mass bins. Description of the columns: (1) simulation tag; (2) subhalo mass range ${\rm log}_{10}({M}_{\rm sub}/(\hMsun))$; (3) median value of the halo mass $\newl(M_{\rm 200}^{\rm crit})$ for centrals; (4) median value of the stellar mass (${\rm log}_{10}({M}_{\rm star}/(\hMsun))$); (5) standard deviation of the stellar mass distribution $\sigma_{\newl  M_{\rm star}} $; (6) median value of halo virial radius $r_{200}^{\rm crit}$ for centrals; (7) median radius within which half of the mass in dark matter is enclosed; (8) median radius within which half of the mass in stars is enclosed; (9) number of haloes; (10) number of {\emph{ satellite} haloes} (11) color used throughout the paper for this particular mass bin and, with different shades, for the simulation from which the mass bin is drawn from.} 
\begin{tabular}{lccrcrrrrrc}
\hline
Simulation tag &  mass bin & $M_{\rm 200}^{\rm crit}$ & $M_{\rm star} $ & $\sigma_{\newl  M_{\rm star}} $ & $r_{200}^{\rm crit}$ & $r_{\rm half}^{\rm dm} $ & $r_{\rm half}^{\rm star} $ & $N_{\rm halo}$ & $N_{\rm sat}$ & Color \\
 & * & * & * & * & ** & ** & ** &  &  \\
 (1)   &(2)            & (3)     & (4)    & (5)    & (6)    & (7)    &(8)  &(9)  &(10) & (11)\\
\hline                                                                                                             
EA L025 & $[10.70 - 11.30]$ & $10.87$ & $8.72$ & $0.46$ & $68.1$ & $28.0$ & $2.3                                      
$ & $234$ & $43$ & black \\                                                                                                
EA L100 & $[11.30 - 12.60]$ & $11.59$ & $9.92$ & $0.45$ & $118.4$ & $50.7$ & $3.2                                     
$ & $4530$ & $745$ & red\\                                                                                              
CO L200 & $[12.60 - 13.70]$ & $12.78$ & $10.88$ & $0.27$ & $295.6$ & $175.7$ & $                                      
31.1$ & $5745$ & $450$ & green\\                                                                                          
CO L400 & $[13.70 - 15.00]$ & $13.82$ & $11.85$ & $0.22$ & $656.3$ & $416.4$ & $                                      
73.5$ & $3014$ & $94$ & blue \\                                                                                           
\hline                                             
\end{tabular}
\label{tbl:sims_stat_IA} 
\end{center}
\vspace{-0.2in}
\footnotetext{* $\newl [M/(\hMsun)]$}
\footnotetext{** $[\hkpc]$}
\end{minipage}
\end{table*}

\subsection{Shape parameter definitions}
\label{Sec:ShapeParameters}
To describe the morphology and orientation of a subhalo we make use of the three-dimensional mass distribution tensor, also referred to as the inertia tensor \citep[ e.g.][]{Cole96},
\begin{equation}\label{eq:inertiatensor}
M_{ij}= \sum_{p=1}^{N_{\rm part}} m_p x_{pi} x_{pj} \, ,
\end{equation}
where $N_{\rm part}$ is the number of particles that belong to the structure of interest, $x_{pi}$ denotes the element $i$ (with $i,j=1,2,3$ for a 3D particle distribution) of the position vector of particle $p$, and $m_p$ is its mass.

 The eigenvalues of the inertia tensor are $\lambda_i$ (with $i=1,2,3$ and $\lambda_1>\lambda_2>\lambda_3$, for a 3D particle distribution as in our case).
The moduli of the major, intermediate, and minor axes of the ellipsoid that have the same mass distribution as the structure of interest, can be written in terms of these eigenvalues as $a=\sqrt{\lambda_1}$, $b=\sqrt{\lambda_2}$, and $c=\sqrt{\lambda_3}$.
Specific ratios of the moduli of the axes are used to define the sphericity, $S=c/a$, and triaxiality, $T= (a^2 - b^2)/(a^2 - c^2)$, parameters \citep[see][]{Velliscig15}.
The eigenvectors $\hat e_i$, associated with the eigenvalues $\lambda_i$, define the orientation of the ellipsoid and are a proxy for the orientation of the structure itself.
We interpret this ellipsoid as an approximation to the shape of the halo and the axis represented by the major eigenvector as the orientation of the halo in a 3D space. 

\section{Orientation-direction alignment}
\label{Sec:Result_orientation_direction}

\begin{figure} \begin{center} 
\includegraphics[width=1.0\columnwidth]{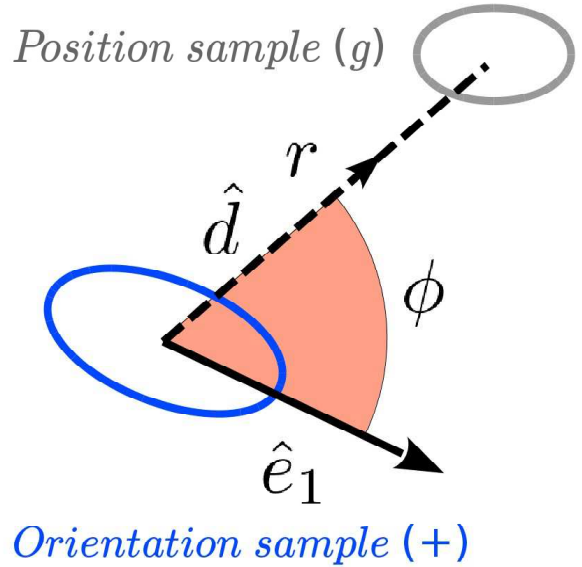} 
 \end{center}
\caption{Diagram of the angle $\phi$ between the major eigenvector $\hat e_1$ of a subhalo in the orientation sample (+), and the separation vector $\hat d$ pointing towards the direction of a subhalo in the position sample (g). Note that all quantities are defined in 3D space. 
}
\label{fig:phi_diagram}
\end{figure}

In this section we present results concerning the alignment between the orientations of the stellar distributions in subhaloes, defined as the major eigenvector of the inertia tensor, $\hat e_1$, and the normalized separation vector, $\hat d$, of a galaxy at distance $r$. Note that all quantities are defined in a 3D space. 
We define $\phi$ as:  
\begin{equation}\label{eq:phi3D}
\phi(r) = \arccos(|\hat e_1 \cdot {\hat d}(r)|) ,
\end{equation}
where $\hat e_1$ is the major eigenvector of a galaxy in the orientation sample, and $\hat d$ is the separation vector pointing towards the position of a galaxy in the position sample (see Fig.~\ref{fig:phi_diagram}). Note that, following Eq.\ref{eq:phi3D}, $0<\phi<\pi/2$.
The value of $\left<\cos(\phi)\right>$ is then computed as an average over pairs of galaxies from the orientation and position samples. Values of  $\left<\cos(\phi)\right>$ close to unity indicate that on average galaxies are preferentially oriented towards the direction of neighbouring subhaloes.
We remind the reader that we use the term subhalo to refer to the ensemble of particles bound to a local minimum in the gravitational potential. Central galaxies are hosted by the most massive subhalo in a FoF group (see \S~\ref{Sec:methods_halodef}).
Throughout the text and in the figures we use (+) to refer to properties of galaxies in the orientation sample, whereas we use (g) for galaxies in the position sample.

Observations typically measure the product of the cosine of the angle $\phi$ and the ellipticity of the galaxy in the orientation sample. We opt to begin our analysis by presenting results only for the angle $\phi$ since it has a clearer interpretation that is independent on the shape determination of the galaxy. We present results for observationally accessible proxies in Section \ref{Sec:obs}.

\subsection{Dependence on subhalo mass and separation}

\label{Sec:auto_cross_mass_cos_phi}
\begin{figure*} \begin{center} \begin{tabular}{cc}
\includegraphics[width=1.0\columnwidth]{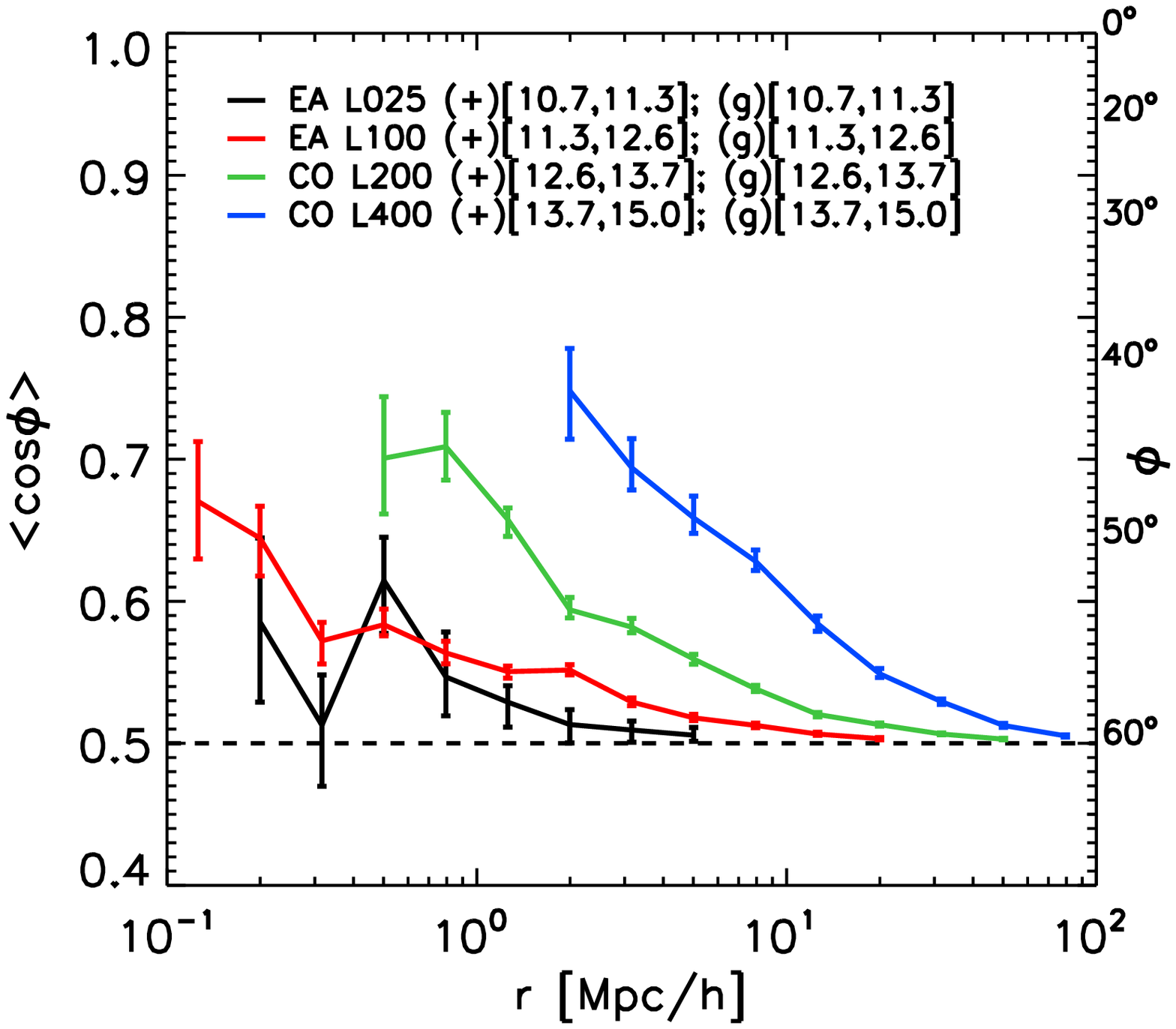} &
\includegraphics[width=1.0\columnwidth]{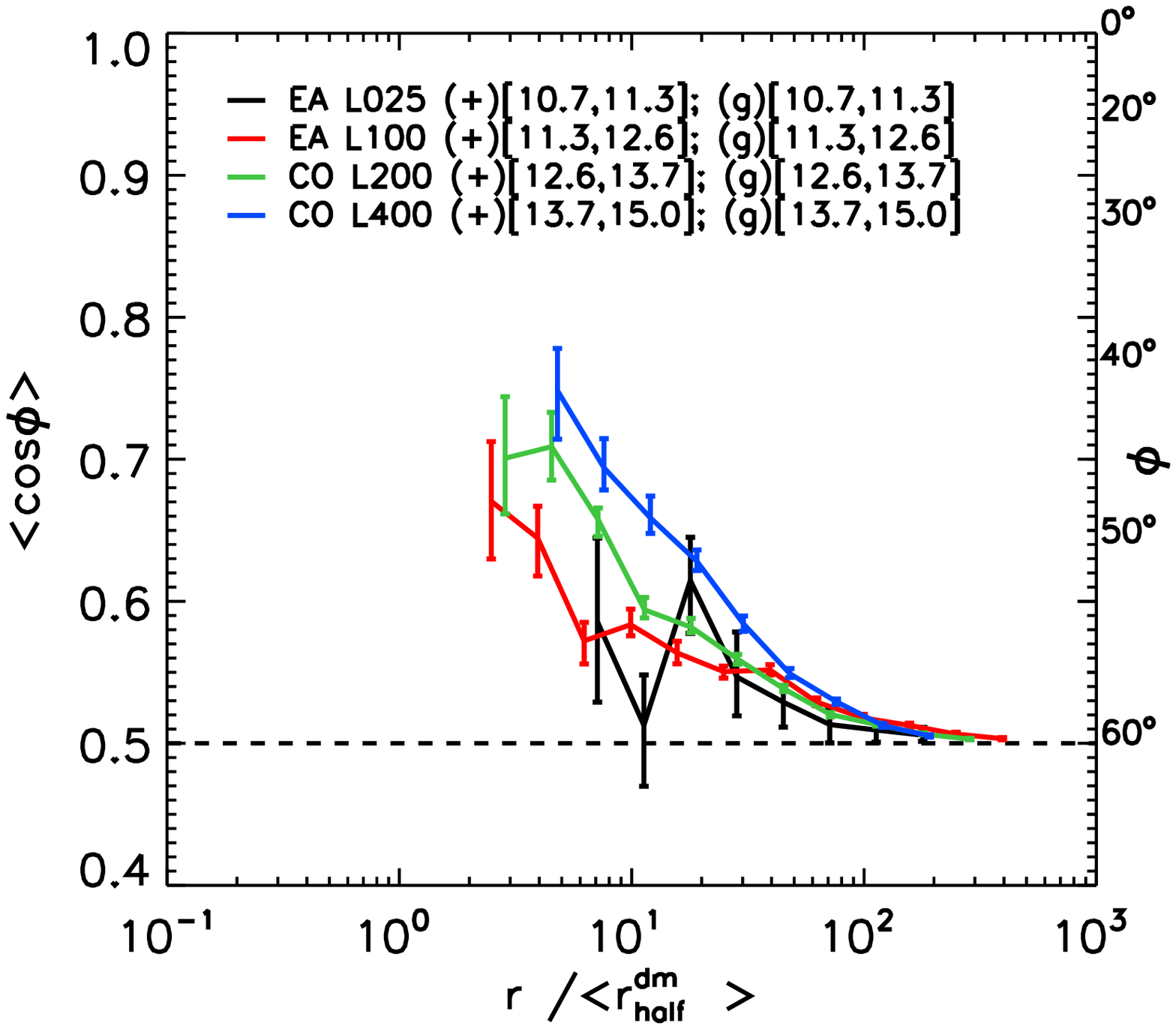}\\
\end{tabular} \end{center}
\caption{ \emph{Left}: Mean value of the cosine of the angle $\phi$ between the major eigenvector of the stellar distribution and the directions towards subhaloes with comparable masses as a function of 3D galaxy separation. Every mass bin is taken from a different simulation. The simulation identifiers used in the legends refer to column (8) of Table~\ref{tbl:simsIA}. The minimum subhalo mass in every bin ensures that only haloes with more than 300 stellar particles are selected. The curves are not shown for 3D separations larger than approximately $1/3$ of the simulation volume. \emph{Right}: Same as left panel but with physical distances rescaled by the $r_{\rm half}^{\rm dm}$ of the subhaloes. In both panels the error bars represent one sigma bootstrap errors. The horizontal dashed line indicates the expectation value for random orientations. The orientation-direction alignment decreases with distance and increases with mass. The mass dependence is greatly reduced when the distances are normalized by $r_{\rm half}^{\rm dm}$.}
\label{fig:total_auto_mass_cos_phi}
\end{figure*} 

\begin{figure*} \begin{center} \begin{tabular}{cc}
\includegraphics[width=1.0\columnwidth]{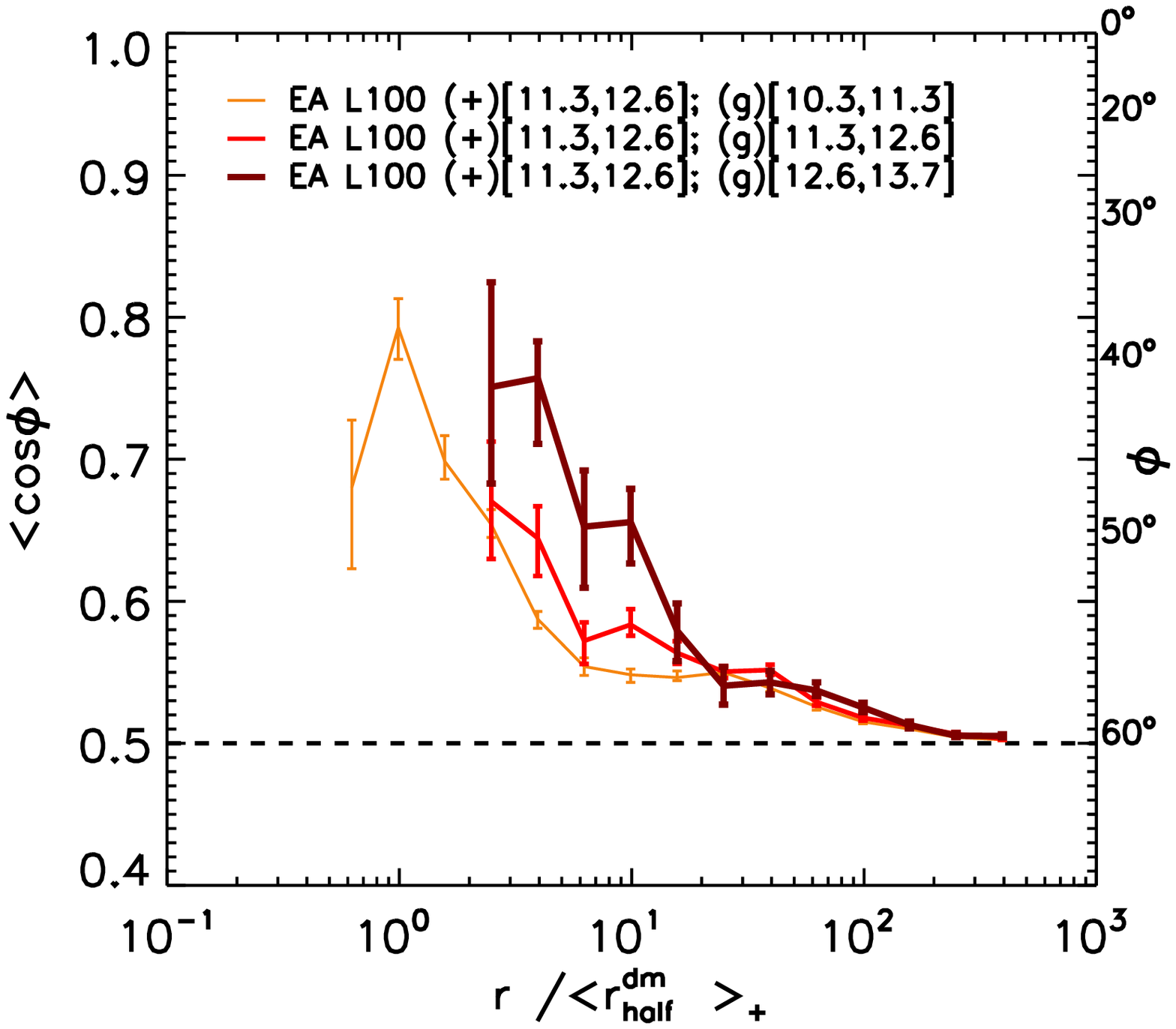} &
\includegraphics[width=1.0\columnwidth]{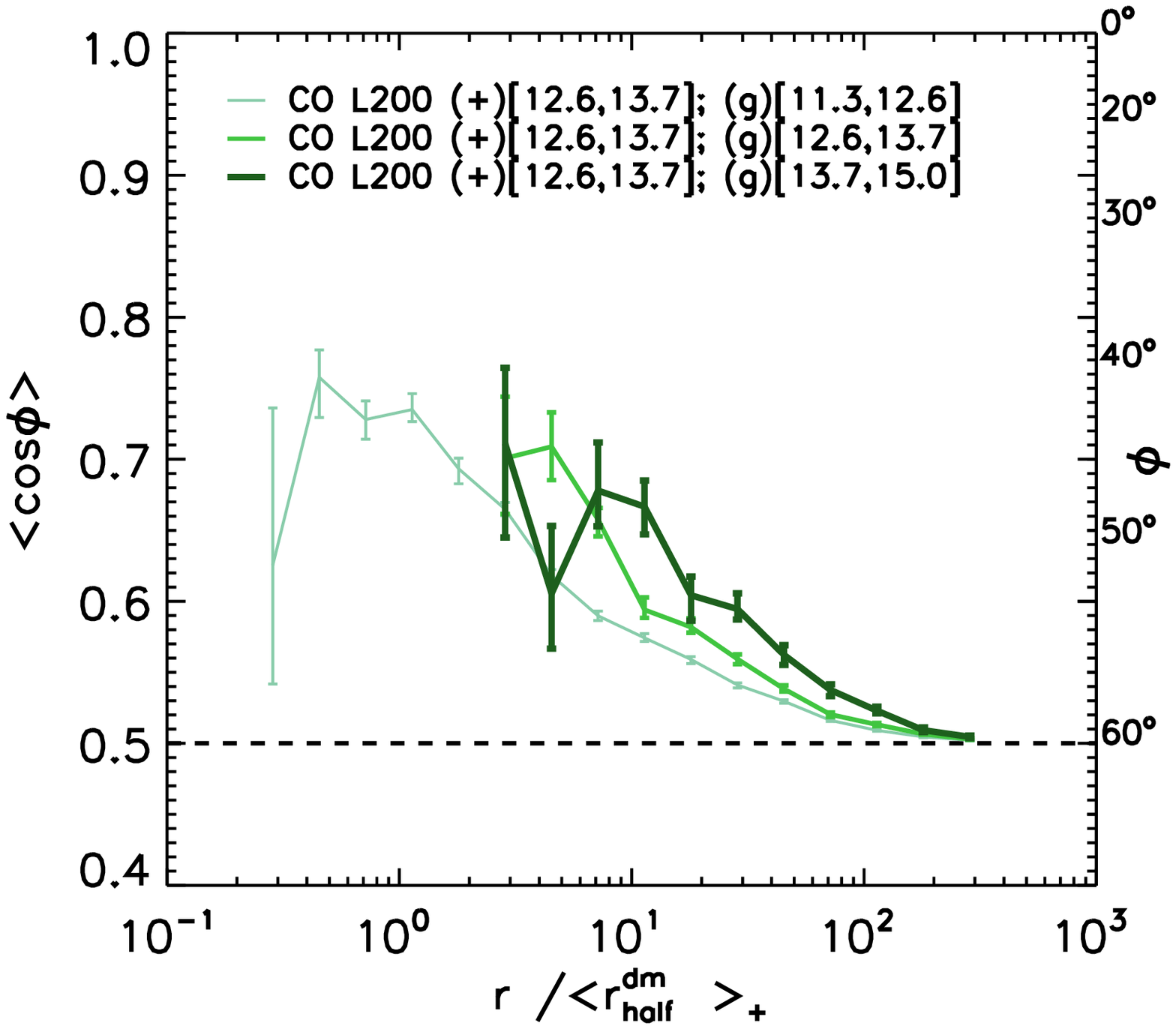}\\
\end{tabular} \end{center}
\caption{ As for the right panel of Fig.~\ref{fig:total_auto_mass_cos_phi}, but in this case the masses of the subhaloes in the orientation sample (+) are kept fixed whereas subhaloes in the position (g) sample are selected from mass bins above, below or equal to the mass bin of the orientation sample. Physical distances are rescaled by the $r_{\rm half}^{\rm dm}$ of the subhaloes in the orientation sample. In the left panel the subhaloes are taken from the EAGLE L100 simulation and in the right panel they are taken from the cosmo-OWLS L200 simulation. In both panels the error bars represent one sigma bootstrap errors. Thicker lines indicate more massive subhaloes for the position sample. The orientation-direction alignment is stronger for more massive subhaloes in the position subsample.}
\label{fig:total_cross_mass_cos_phi}
\end{figure*} 

The left panel of Fig.~\ref{fig:total_auto_mass_cos_phi} shows $\left<\cos(\phi)\right>$ for pairs of galaxies (both centrals and satellites) binned in subhalo mass and as a function of 3D separation $r$. 
Subhaloes in the orientation sample (+) are chosen to have the same mass limits as the subhaloes in the position sample (g).
Values are shown for four different choices of subhalo masses, where every mass bin is taken from a different simulation (see legend). Errors are estimated via the bootstrap technique. Specifically, we use the 16th and the 84th percentiles of 100 realizations to estimate the lower and upper limits of the error bars.
The cosine of the angle between the orientation of galaxies and the direction of neighbouring galaxies is a decreasing function of distance and it increases with mass. For large separations the angle tends to the mean value for a randomly distributed galaxy orientation, i.e. $\left<\cos(\phi)\right>=0.5$. 
The physical scale at which this asymptotic behaviour is reached increases with increasing subhalo mass.

The right panel of Fig.~\ref{fig:total_auto_mass_cos_phi} shows $\left<\cos(\phi)\right>$ as a function of the physical separation rescaled by the average size of subhaloes, $\left<r_{\rm half}^{\rm dm}\right>$, in that mass bin. This rescaling 
removes most, but not all, of the offset between different halo mass bins.
On average, subhalo pairs separated by more than $100r_{\rm half}^{\rm dm}$ show only weak alignment ($\left<\cos(\phi)\right> \leq 0.52$ at $100r_{\rm half}^{\rm dm}$).

Fig.~\ref{fig:total_cross_mass_cos_phi} shows $\left<\cos(\phi)\right>$ as a function of the separation rescaled by the average size of the subhaloes in the orientation (+) sample. In this case the masses of the subhaloes for which we measure the orientation of the stellar distribution are kept fixed whereas haloes in the position (g) sample are selected from mass bins above, below or equal to the mass bin of the orientation sample. 
Results are shown for two of the four simulations: in the left panel for EAGLE L100 and in the right panel for cosmo-OWLS L200. The line thickness is proportional to the subhalo mass of the position sample. 
The orientation of subhaloes of a given mass tends to be more aligned with the position of higher-mass subhaloes.

We note that the two suites of simulations employed here, cosmo-OWLS and EAGLE, differ in resolution, volume, cosmology and subgrid physics. Testing how each of these differences impacts our mean results is beyond the scope of this study (we would need as many simulations as differences that we wish to test), therefore we examine the overall convergence of the two simulations by selecting a subhalo mass bin $12.6 < \newl(M_{\rm sub}/ [\hMsun]) < 13.1$ that yields an orientation sample of galaxies that is numerous enough in the EAGLE L100 simulation, as well as resolved in the cosmo-OWLS L200 simulation. We find that, in this specific case, the results are consistent within the bootstrapped errors, both for stars and stars within $r^{\rm star}_{\rm half}$ (not shown).

Subhalo mass  plays an important role in the strength of the orientation-direction alignment of subhaloes.
The dependence on the subhalo mass weakens with distance but only becomes negligible for separation $\gg100$ times the subhalo radius. 

\subsection{Dependence on the choice of matter component}
\label{Sec:total_dm_star_halfstar_cos_phi}

\begin{figure*} \begin{center} \begin{tabular}{cc}
\includegraphics[width=1.0\columnwidth]{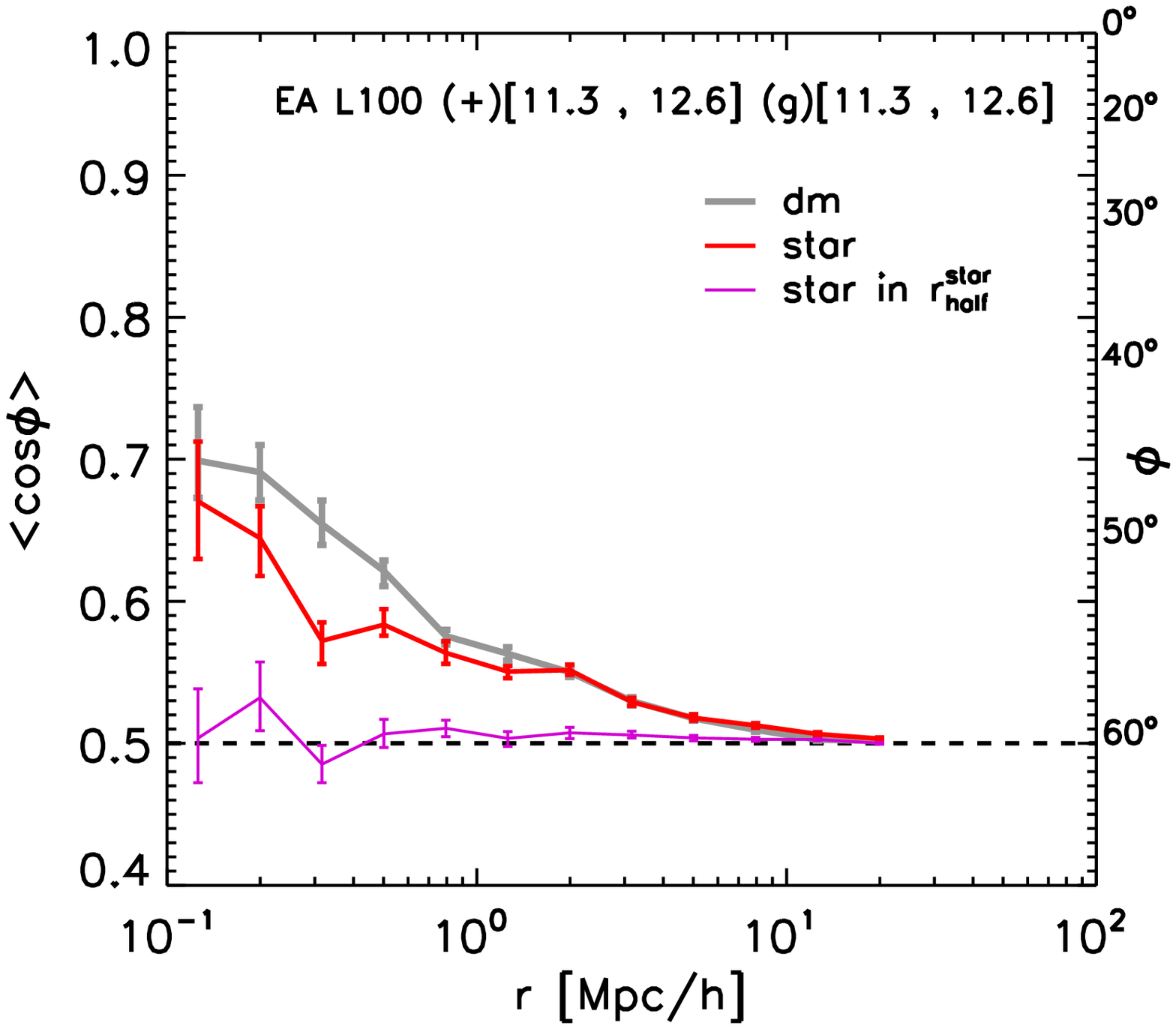} &
\includegraphics[width=1.0\columnwidth]{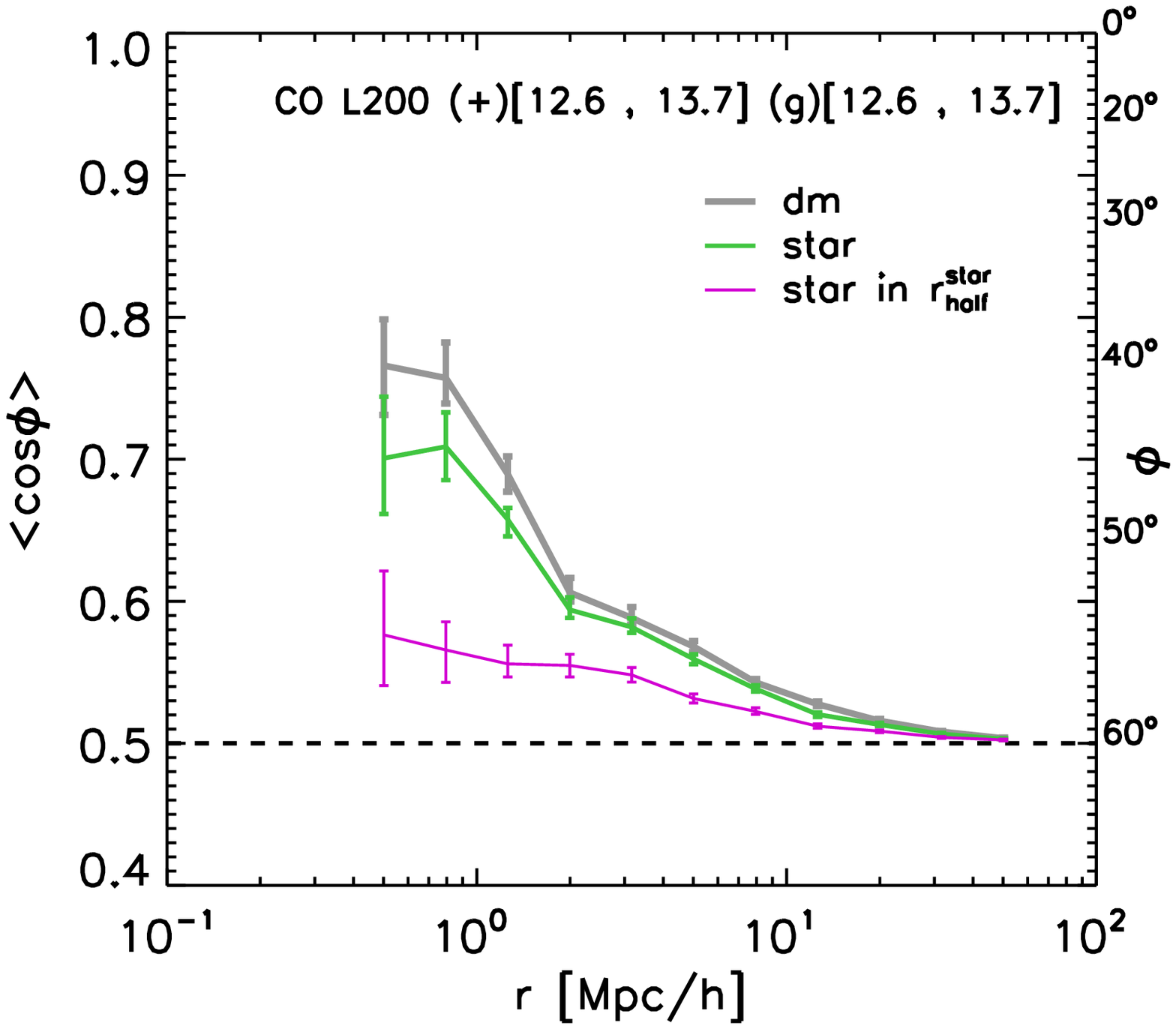}\\
\end{tabular} \end{center}
\caption{ Mean value of the cosine of the angle $\phi$ between the major eigenvectors of the distributions of stars (red curve in the left panel and green curve in right panel as in Fig.~\ref{fig:total_auto_mass_cos_phi}), dark matter (gray curves), or stars within $r_{\rm half}^{\rm star}$ (purple curves) and the direction towards subhaloes with comparable masses as a function of 3D galaxy separation. The subhaloes used for the left panel are taken from the EAGLE L100 ($11.3 < \newl(M_{\rm sub}/ [\hMsun]) < 12.6$) simulation while in the right panel they are taken from the cosmo-OWLS L200 simulation ($12.6 < \newl(M_{\rm sub}/ [\hMsun]) < 13.7$). Thicker lines indicate components with stronger alignment. In both panels the error bars represent one sigma bootstrap errors. The orientation of the dark matter component is most strongly aligned with the directions of nearby subhaloes, whereas the orientation of stars inside $r^{\rm star}_{\rm half}$ shows the weakest alignment.}
\label{fig:total_dm_star_halfstar_cos_phi}
\end{figure*} 

In this section we report the orientation-direction alignment for the case in which the orientation  of the subhalo is calculated using, respectively, dark matter, stars (as in the previous section) and stars within the half-mass radius $r^{\rm star}_{\rm half}$. An alternative choice of a proxy for the typical extent of a galaxy would be to consider only stars within a fixed 3D aperture of 30Kpc that gives similar galaxy properties as the 2-D Petrosian apertures often used in observational studies \citep[][]{Schaye14}. Note that the two definitions coincide for the subhalo mass bin $12.6 < \newl(M_{\rm sub}/ [\hMsun]) < 13.7$ (CO L200). We note that $r^{\rm star}_{\rm half}$ varies among the four mass bins used in this work (see Table~\ref{tbl:sims_stat_IA} column (8)).

Fig.~\ref{fig:total_dm_star_halfstar_cos_phi} shows the cosine of the angle $\phi$ between the direction of nearby subhaloes and the orientation of the distribution of dark matter, stars (as shown in Fig.~\ref{fig:total_auto_mass_cos_phi}) and stars within the half-mass radius of subhaloes in the same mass bin.
The left panel displays the results for the subhalo mass bin $11.3 < \newl(M_{\rm sub}/ [\hMsun]) < 12.6$ (from the EAGLE L100 simulation), whereas the right panel  refers to the subhalo mass bin $12.6 < \newl(M_{\rm sub}/ [\hMsun]) < 13.7$ (from the cosmo-OWLS L200 simulation).

Irrespective of the subhalo mass and separation, the orientation of the dark matter component shows the strongest alignment with the directions of nearby haloes, whereas the orientation of stars inside $r^{\rm star}_{\rm half}$ shows the weakest alignment.

These results are suggestive of a scenario in which the alignment between subhaloes and the surrounding density field is imprinted mostly on the dark matter distribution.
Therefore, when the orientation of the subhalo is computed using all stars or the stars within $r^{\rm star}_{\rm half}$, the signal is weakened according to the internal misalignment angle between the specified component and the total dark matter distribution. 
The trend shown by Fig.~\ref{fig:total_dm_star_halfstar_cos_phi} therefore follows  naturally from the results of \citet{Velliscig15}:
stars within $r^{\rm star}_{\rm half}$ exhibit a weaker alignment with the total dark matter distribution than all stars in the subhalo.

The difference between the orientation-direction alignment obtained using the dark matter, all the stars or the stars within the typical extent of the galaxy, could account for the common finding reported in the literature of galaxy alignment, that such alignments are systematically stronger in simulations 
than when measured in observational data (see the recent reviews of \citealt{Kiessling15} and \citealt{Kirk15} for a detailed comparison between observational and computational studies).
Observations are limited to the shape and orientation of the region of a galaxy above a limit surface brightness (often within surface brightness isophotes), whereas simulations need to rely on proxies for the extent of those regions \citep[e.g. using baryonic overdensity thresholds][]{Hahn10, Codis15, Welker14, Dubois14} or to employ weighting schemes to the sample of star particles that constitute a galaxy \citep[see e.g. use of the \emph{reduced} inertia tensor in][]{Tenneti14b}.

\subsection{Dependence on galaxy morphology}
\label{Sec:morpho_cos_phi}

\begin{figure*} \begin{center} \begin{tabular}{cc}
\includegraphics[width=1.0\columnwidth]{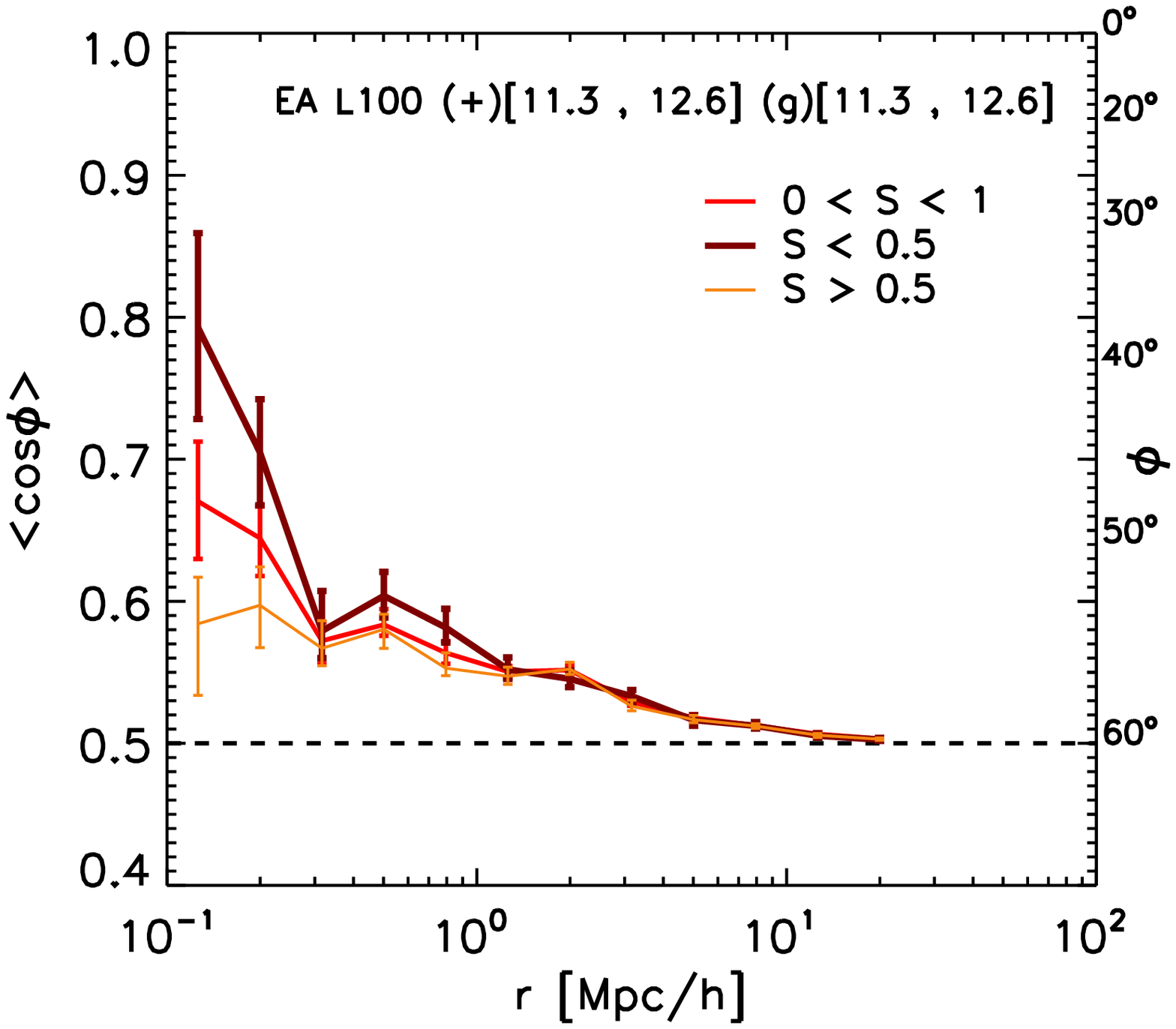} &
\includegraphics[width=1.0\columnwidth]{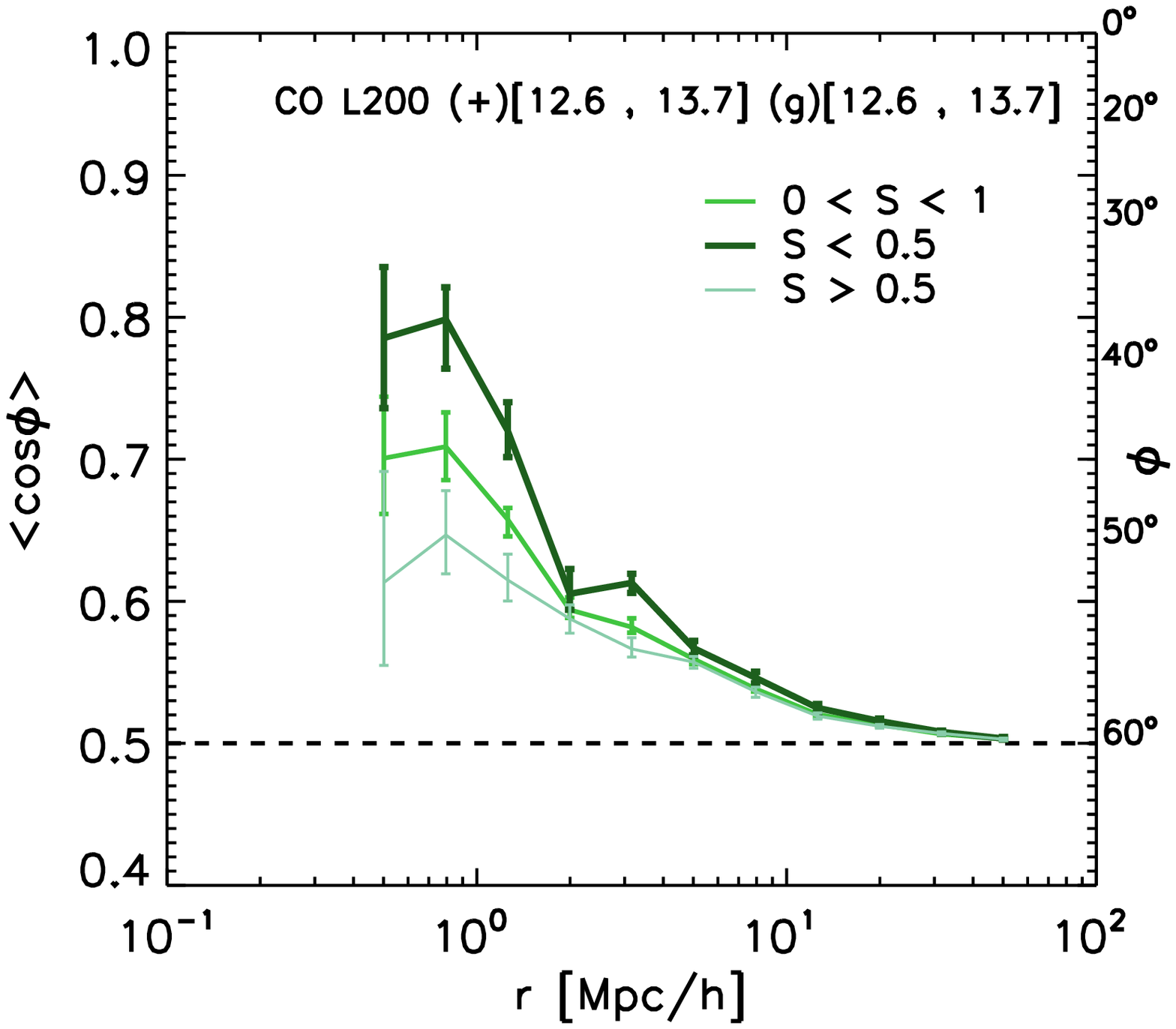}\\
\end{tabular} \end{center}
\caption{Mean value of the cosine of the angle $\phi$ between the major eigenvector of the stellar distribution and the direction towards neighbouring subhaloes as a function of 3D galaxy separation, for galaxies in the orientation sample selected based on their shape. The selection is based on the sphericity of the whole stellar distribution defined as $S=c/a$ where $a$ and $c$ are the square root of the major and minor eigenvalues of the inertia tensor respectively. We choose a threshold value for the sphericity of $0.5$.  The subhaloes used for the left panel are taken from the EAGLE L100 ($11.3 < \newl(M_{\rm sub}/ [\hMsun]) < 12.6$) simulation while in the right panel they are taken from the cosmo-OWLS L200 simulation ($12.6 < \newl(M_{\rm sub}/ [\hMsun]) < 13.7$). Thicker lines indicate components with stronger alignment. In both panels the error bars represent one sigma bootstrap errors. More spherical galaxies show a weaker orientation-direction alignment.}
\label{fig:morpho_cos_phi}
\end{figure*}

Theory predicts that the alignment of early-type galaxies  and late-type galaxies arises from different physical processes \citep[e.g.][]{Catelan00}. 
It is of interest then to study the alignment as a function of galaxy morphologies.

In this section we report the orientation-direction alignment of galaxies with different sphericities in order to explore the effect of the shape of galaxies on the orientation-direction alignment.
We divide our sample of subhaloes according to the sphericity of their whole stellar distribution, defined as $S=c/a$ where $a$ and $c$ are the squareroot of the major and minor eigenvalues of the inertia tensor respectively (see \S\ref{Sec:ShapeParameters}). We choose a threshold value for the sphericity of $0.5$ that yields a similar numbers of galaxies in the two subsamples, as the median sphericity of the total sample is $0.55$. This galaxy selection by sphericity represent a simple proxy for galaxy morphology.

Fig.~\ref{fig:morpho_cos_phi} shows the mean values of the cosine of the angle $\phi$ for galaxies of sphericity above and below the threshold, as well as for the total sample. The left panel displays the results for the subhalo mass bin $11.3 < \newl(M_{\rm sub}/ [\hMsun]) < 12.6$ (from the EAGLE L100 simulation), whereas the right panel  refers to the subhalo mass bin $12.6 < \newl(M_{\rm sub}/ [\hMsun]) < 13.7$ (from the cosmo-OWLS L200 simulation).

More spherical galaxies (thinner lines) show a weaker orientation-direction alignment. 
The differences between the two shape selected samples of haloes are within the errors for scales larger than $1\hMpc$, suggesting that the effect of shape is dominated by subhaloes of the same hosts.
A similar trend (not shown) is found using  triaxiality, see \S\ref{Sec:ShapeParameters}, as the indicator of galaxy shape. Prolate ($T>0.5$)  stellar distributions show the strongest orientation-direction alignment, whereas oblate ($T<0.5$) ones show the weakest.
The better alignment of prolate or aspherical galaxies is probably due to the fact that these galaxies align better with their underlying  dark matter distributions (not shown), which in turn produces a stronger orientation-direction alignment (see Fig.~\ref{fig:total_dm_star_halfstar_cos_phi}).

We note that the orientation of a perfectly spherical  distribution ($S=1$) of stars is ill defined. Although this can potentially affect our measurements, less than 2\% of galaxies in our sample have a sphericity higher than 0.8.
We also note that more massive haloes, for which the orientation-direction alignment is strongest, tend to be less spherical and more triaxial \citep[see][]{Velliscig15}. Therefore, selecting haloes by shape biases the sample towards systematically different masses: however, the mass difference in the two shape-selected samples is about 4\%, which is too small to explain the differences in alignment of haloes with different shapes.

Observations indicate that ellipsoidal galaxies show stronger intrinsic alignment than blue disk galaxies \citep[][]{Hirata07, Singh14}.
However, we caution the reader that there are still many complications to take into account before one can compare the trends discussed above with these observational results.
First, one would need to select galaxies based on their colors, which requires stellar population synthesis models. Second, the sphericity of the stellar component is a simplistic proxy for selecting disc galaxies. 
Selecting galaxies according to their morphology, in a similar way as done observationally, would require a stellar light decomposition in bulge and disc component.

\subsection{Alignment of satellite and central galaxies}
\label{Sec:CenSat}
\subsubsection{The increased probability of finding satellites along the major axis of the central galaxy}
\label{CenSat_cos_phi}

\begin{figure*} \begin{center} \begin{tabular}{cc}
\includegraphics[width=1.0\columnwidth]{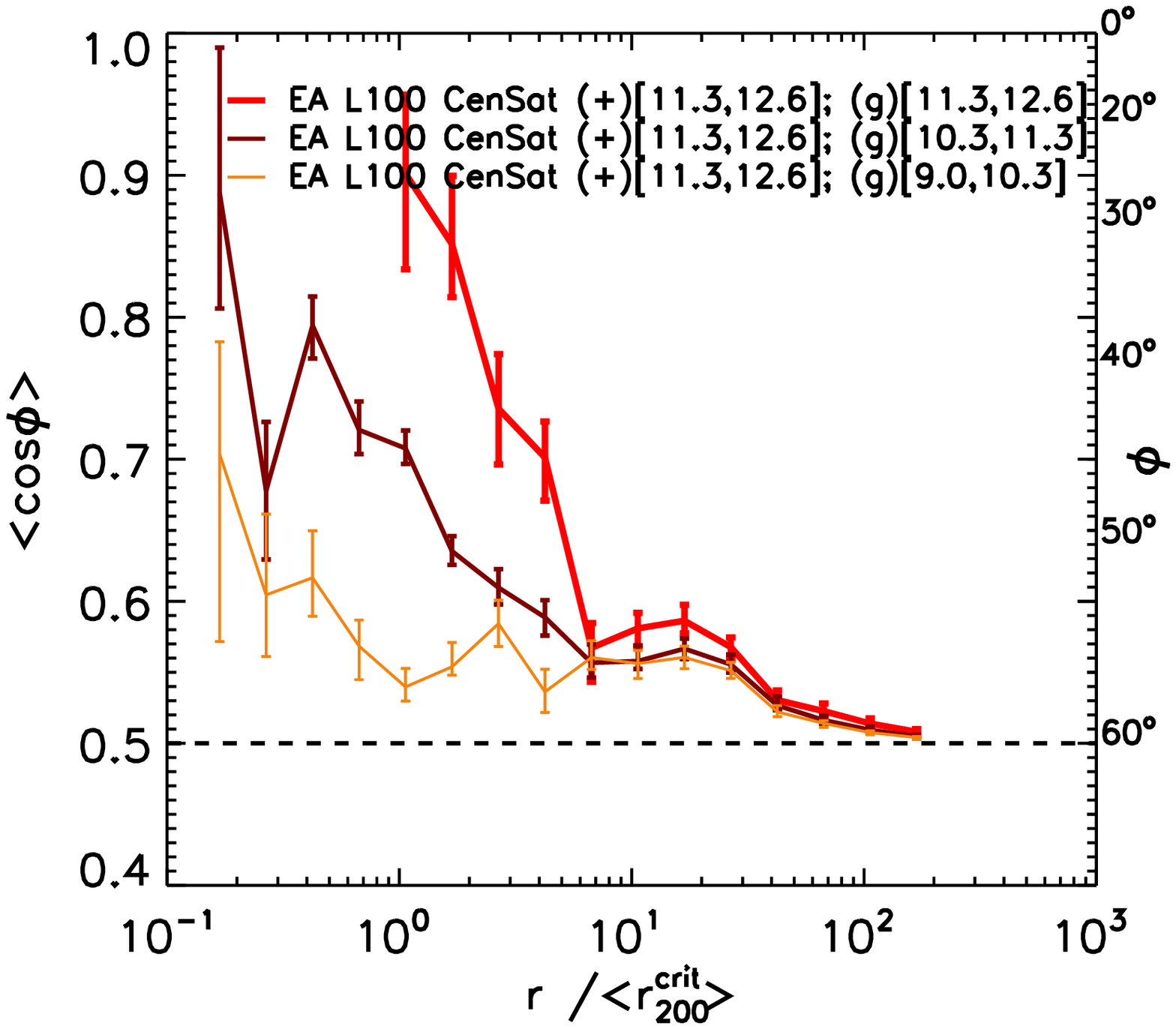} &
\includegraphics[width=1.0\columnwidth]{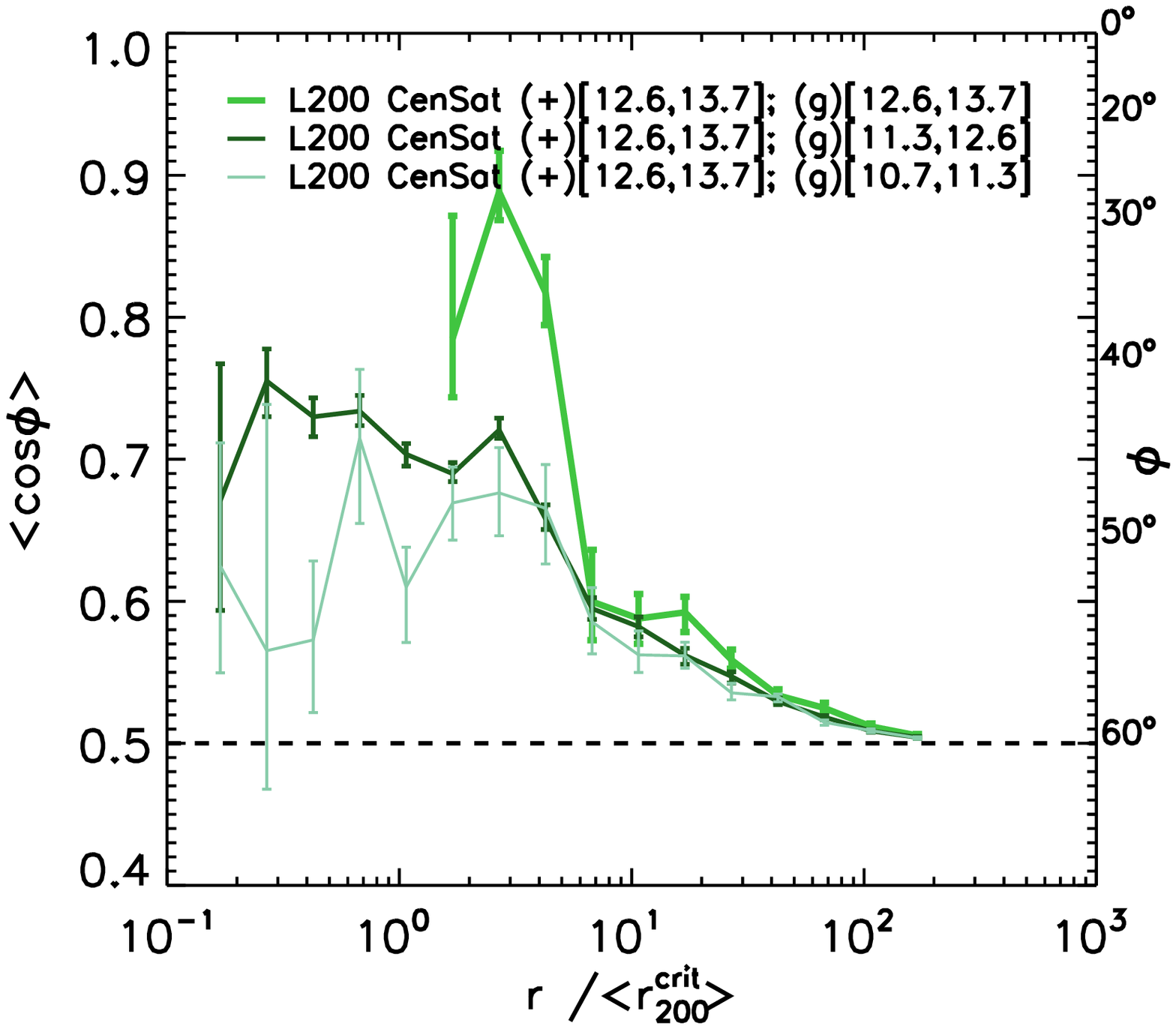}\\
\end{tabular} \end{center}
\caption{ Mean value of the cosine of the angle between the orientation of the stars in the central galaxy and the direction of satellite galaxies as a function of the 3D galaxy separation, rescaled by the host halo $r^{\rm crit}_{200}$. The central galaxies used for the left panel are taken from the EAGLE L100  ($11.3 < \newl(M_{\rm sub}/ [\hMsun]) < 12.6$) while in the right panel they are taken from the cosmo-OWLS L200 simulation ($12.6 < \newl(M_{\rm sub}/ [\hMsun]) < 13.7$). In both panels the error bars represent one sigma bootstrap errors. Thicker lines indicate higher mass. The satellite distribution is aligned with the central galaxy out to $\sim 100r^{\rm crit}_{200}$. For $r< 10r^{\rm crit}_{200}$ the alignment is substantially stronger for higher-mass satellites.}
\label{fig:CenSat_cos_phi}
\end{figure*} 

\begin{figure*} \begin{center} \begin{tabular}{cc}
\includegraphics[width=1.0\columnwidth]{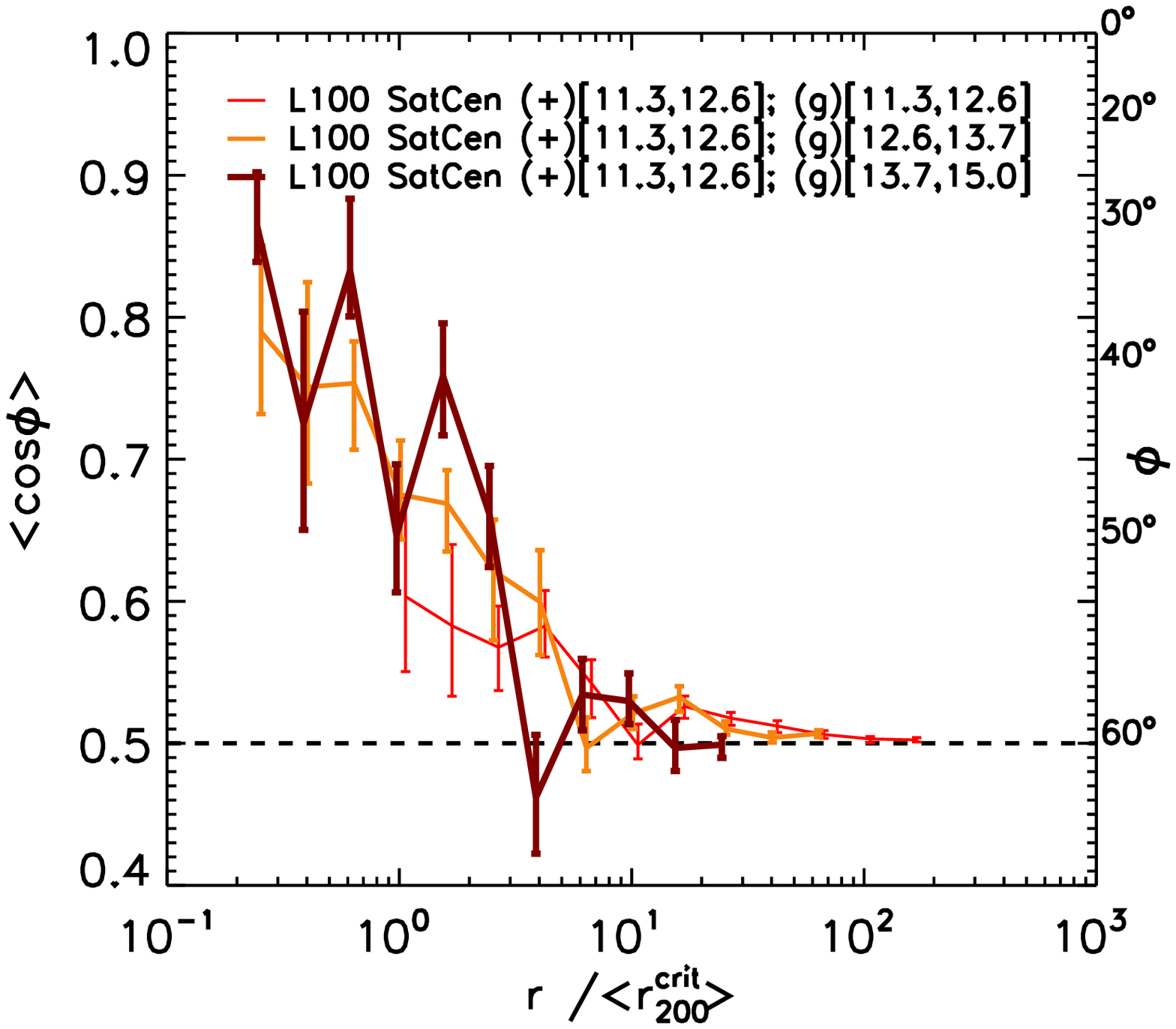} &
\includegraphics[width=1.0\columnwidth]{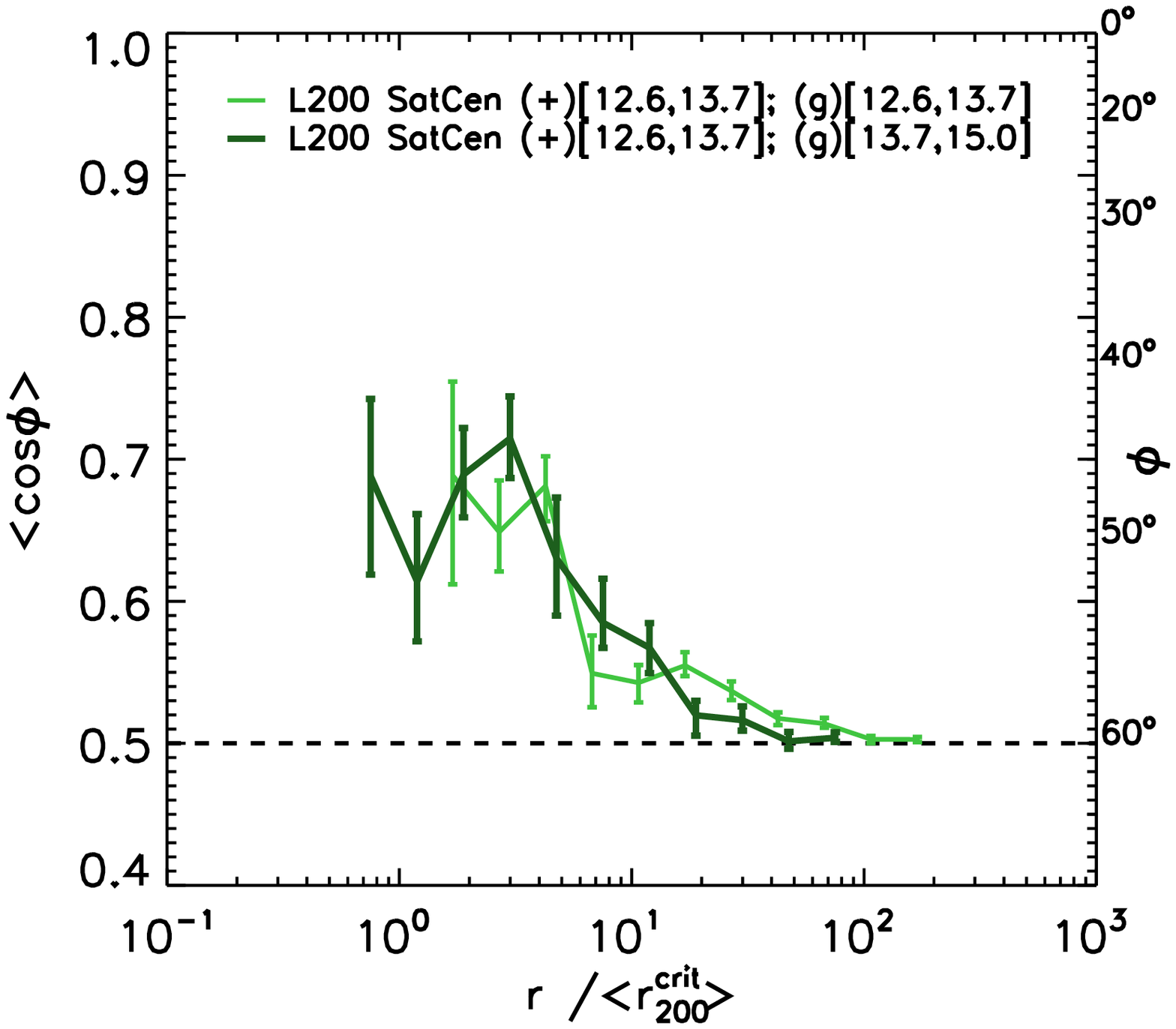}\\
\end{tabular} \end{center}
\caption{As for Fig.~\ref{fig:CenSat_cos_phi} but in this case the orientation is computed for the stellar distribution in satellite galaxies and the angle is measured with respect to the directions of central galaxies hosted by subhaloes of different masses. The alignment between satellites and the directions of centrals decreases with distance but is insensitive to the mass of the host halo. }
\label{fig:SatCen_cos_phi}
\end{figure*} 

In the previous sections we studied the orientation-direction alignment of galaxies irrespective of their classification as centrals or satellites.
In this subsection we report the alignment between the orientations of central galaxies (g) and the directions of  satellite\footnote{In this subsection, satellite galaxies do not necessarily belong to the same haloes that host the paired central galaxies.} galaxies (+)
and, in turn, the probability of finding satellite galaxies distributed along the major axis of the central galaxy.
This effect has been studied both theoretically, making use of N-Body \citep[e.g.][]{Faltenbacher08, Agustsson10, Wang14a} and hydrodynamical simulations \citep[][]{Libeskind07, Deason11}, and observationally \citep[][]{Sales04, Brainerd05, Yang06, Wang08, Nierenberg12, Wang14, Dong14}. Those studies report that the distribution of satellites around central galaxies is anisotropic, with an excess of satellites aligned with the major axis of the central galaxy.

Fig.~\ref{fig:CenSat_cos_phi} shows the average angle between the orientation of the stellar distribution of central subhaloes and the position of satellite galaxies. Values of  $\left< \cos\phi \right>$ that are significantly greater than $0.5$ indicate that the positions of satellites are preferentially aligned with the major axis of the central galaxy. We use two different mass bins taken from two simulations: $11.3 < \newl(M_{\rm sub}/ [\hMsun]) < 12.6$ from EAGLE L100 (left) and  $12.6 < \newl(M_{\rm sub}/ [\hMsun]) < 13.7$ from cosmo-OWLS L200 (right). The line thickness is proportional to the subhalo mass of the position (g) sample. 
In both panels the physical separations between the pairs are normalized by the $\left<r^{\rm crit}_{200}\right>$ of the haloes hosting the central galaxies.

For separations up to $100\left<r^{\rm crit}_{200}\right>$, the positions of satellite galaxies are significantly aligned with the orientation of  central galaxies (not necessarily in the same host halo), with more massive satellites showing a stronger alignment. 
The same qualitative behaviour is found for both mass bins, but the effect is stronger for the more massive central subhaloes.
On scales larger than $\sim \left<10 \, r^{\rm crit}_{200}\right>$ the alignment depends only weakly on the mass of the satellite subhaloes.  
We speculate that the alignment of satellites with central galaxies of different host haloes is likely driven by the correlation between the orientation of the central galaxies and the surrounding large-scale structure, which in turn influences the positions of satellite galaxies.

\subsubsection{The radial alignment of satellite galaxies with the direction of the host galaxy}
\label{Sec:SatCen_cos_phi}

\begin{figure*} \begin{center} \begin{tabular}{cc}
\includegraphics[width=1.0\columnwidth]{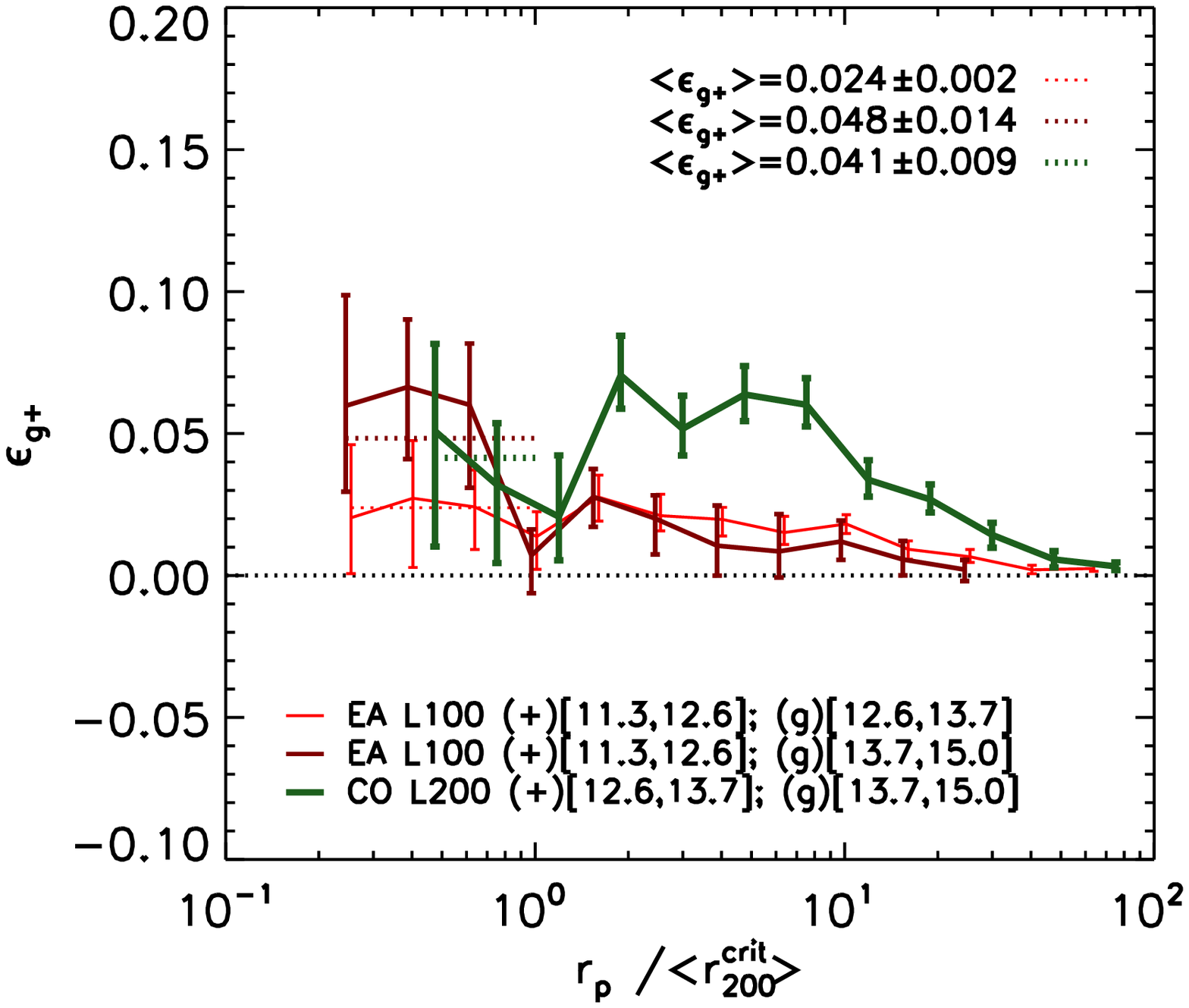} &
\includegraphics[width=1.0\columnwidth]{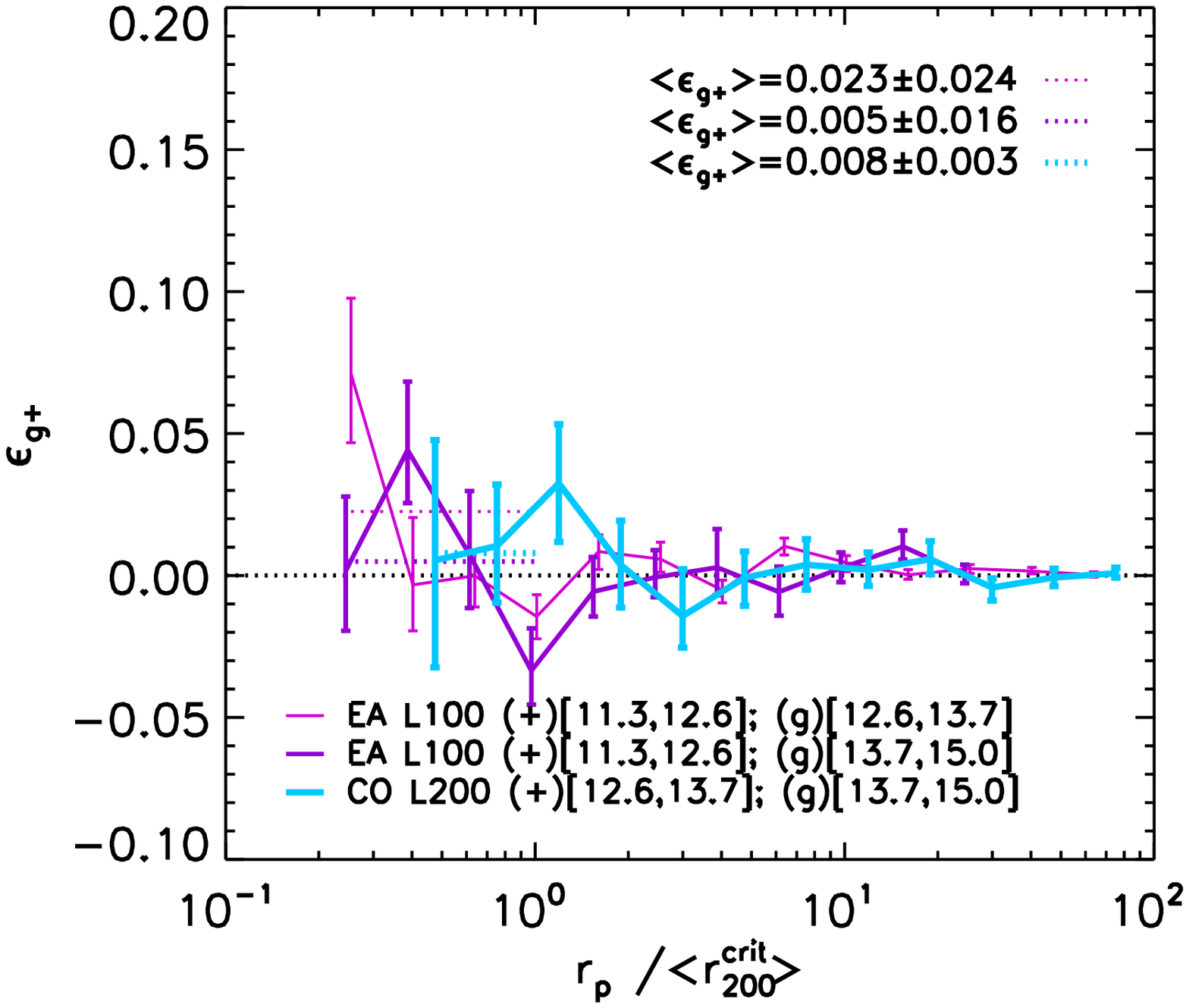}\\
\end{tabular} \end{center}
\caption{Values of the observationally accessible proxy for orientation-direction alignment, $\epsilon_{\rm g+}$  (see Eq.~\ref{eq:epsilon}), as a function of the projected separation $r_{\rm p}$. Only pairs that are separated by less than  $2.5\hMpc$ along the projected axis are considered for this analysis. The error bars indicate one sigma bootstrap errors. In the left panel the orientation of the subhalo is measured using all the stellar particles whereas in the right panel only  stars  within the $r_{\rm half}^{\rm star}$ are used, which greatly reduces the alignment. Thicker lines indicates higher masses. The coloured dotted lines show the average value of $\epsilon_{\rm g+}$ within the virial radius of the central galaxy. Observational measurements from \citet{Sifon15} constrained the average ellipticity to be $\epsilon_{\rm g+}= - 0.0037 \pm 0.0027$.}
\label{fig:groups_cluster_scales_epsilon_g_plus}
\end{figure*}

Here we investigate the radial alignment of the orientations of satellites (+) with the direction of the central galaxy (g), whereas in the previous section report the results for alignment between the orientations of the central galaxy and the direction of satellites. The orientation of satellite subhaloes is computed using all the stars bounded to the subhalo. Theoretical studies using N-body simulations \citep[][]{Kuhlen07, Pereira08, Faltenbacher08} and hydrodynamic simulations \citep[][]{Knebe10} found that on average the orientation of satellite galaxies is aligned with the direction of the centre of their host halo. 

Fig.~\ref{fig:SatCen_cos_phi} shows the average value of the cosine of the angle between the orientation of the satellite and the direction of the centrals as a function of the separation rescaled by the average virial radius ($r_{200}^{\rm crit}$). The mass of the subhaloes in the orientation sample (+) is kept fixed whereas the masses of the central haloes (g) are chosen to have similar or higher masses. Values of  $\left< \cos\phi \right>$ that are significantly greater than $0.5$ indicate that the orientation of satellites galaxies  are preferentially aligned  towards the direction of central galaxies. As for the previous subsection, we use two different mass bins taken from two simulations: $11.3 < \newl(M_{\rm sub}/ [\hMsun]) < 12.6$ from EAGLE L100 (left) and  $12.6 < \newl(M_{\rm sub}/ [\hMsun]) < 13.7$ from cosmo-OWLS L200 (right). The line thickness is proportional to the subhalo mass of the position (g) sample. 
In both panels the physical separations between the pairs are normalised by the $\left<r^{\rm crit}_{200}\right>$ of the haloes hosting the central galaxies.

The major axes of satellite galaxies, when all stars are considered, are significantly aligned towards the direction of the centrals within their virial radius. The strength of the alignment declines very rapidly with radius and is very small outside the virial radius. There is only a weak dependence on the central subhalo mass.

We note that by considering only stars in $r_{\rm half}^{\rm star}$ the trends shown in Fig.~\ref{fig:CenSat_cos_phi} and in Fig.~\ref{fig:SatCen_cos_phi} are weakened (not shown). This results in a less significant alignment for galaxies hosted by subhaloes with masses $11.3 < \newl(M_{\rm sub}/ [\hMsun]) < 12.6$ from the EAGLE L100 simulation, whereas a still significant alignment is found for galaxies with $12.6 < \newl(M_{\rm sub}/ [\hMsun]) < 13.7$ from the cosmo-OWLS L200 simulation.

\section{Towards observations of orientation-direction galaxy alignment}
\label{Sec:obs}

In this subsection we report results for observationally accessible proxies for the orientation-direction alignment, which depend on the shape of galaxies as well as on their orientation, making them tightly connected to cosmic shear studies. All the relevant quantities for the following analysis are defined in a 2D space. 

Observationally, the ellipticity is decomposed into the projected tangential ($\epsilon_+$) and transverse ($\epsilon_{\times}$) components with respect to the projected separation vector of the galaxy pair:
\begin{align}
\epsilon_+ & = |\epsilon|\cos(2\Phi) \label{eq:epsilon_plus}\\
\epsilon_{\times} & = |\epsilon|\sin(2\Phi) \\
|\epsilon| & = \frac{1 - b/a}{1 + b/a} \, ,
\end{align}
where $\Phi$ is the position angle\footnote{The symbol $\Phi$ is used to indicate an angle between vector in 2D, whereas the symbol $\phi$ (see Eq.~\ref{eq:phi3D}) indicates an angle between vectors in 3D.}
between the projected orientation of the galaxy and the direction of a galaxy at projected distance $r_{\rm p}$ and $b/a$ is the axis ratio of the projected galaxy.

Then the function $\epsilon_{\rm g+}$ is defined as:
\begin{equation}
\label{eq:epsilon}
\epsilon_{\rm g+}(r_p)=  \sum_{i\neq j\mid r_{p}}\frac{\epsilon_{+}(j\mid i)}{N_{\rm pairs}},
\end{equation}
where the index $i$ represents a galaxy in the shape sample, whereas the index $j$ represents a galaxy in the position sample.
The function $\epsilon_{\rm g+}(r_{\rm p})$ is the average value of $\epsilon_{+}$ at the projected separation $r_{\rm p}$.

Groups and clusters of galaxies, where strong tidal torques are expected to align satellite galaxies toward the centre of the host's gravitational potential, are ideal environments to study orientation-direction alignment. However, the task of measuring this alignment has proven to be very challenging \citep[see][and references therein]{Kirk15}.  In group and cluster environments, the measured quantity, $\epsilon_{\rm g+}$ (see Eq.~\ref{eq:epsilon}), is the mean value of the angle between the projected orientation of the \emph{satellite} galaxy and the direction of the \emph{host}, multiplied by the projected ellipticity of the satellite.
Typical values of the root mean square of galaxy shape parameter, $e=(1-(b/a)^2) / (1+(b/a)^2)$, in 
the set of simulations employed in this study, can be found in Fig.~5 of \citet{Velliscig15}. 
Those values are in broad agreement with the observed noise-corrected values (about $0.5$-$ 0.6$ depending on luminosity and galaxy type, \citep[e.g.][]{Joachimi13})
when all stars in subhaloes are considered. However, when only stars within $r_{\rm half}^{\rm star}$ are considered, \citet{Velliscig15} found typical values for $e_{\rm rms}$ of $\approx 0.2$-$ 0.3$, that is a factor of 2 lower than the observed value. This suggests that  galaxy shapes computed using stars within $r_{\rm half}^{\rm star}$ are rounder than the observed shapes, potentially leading to an underestimate of the $\epsilon_{\rm g+}$. To quantify this effect, we would need to analyse synthetic galaxy images from simulations with the shape estimator algorithms used in weak lensing measurements. We defer such an investigation to future works.

\begin{figure} \begin{center} 
\includegraphics[width=1.0\columnwidth]{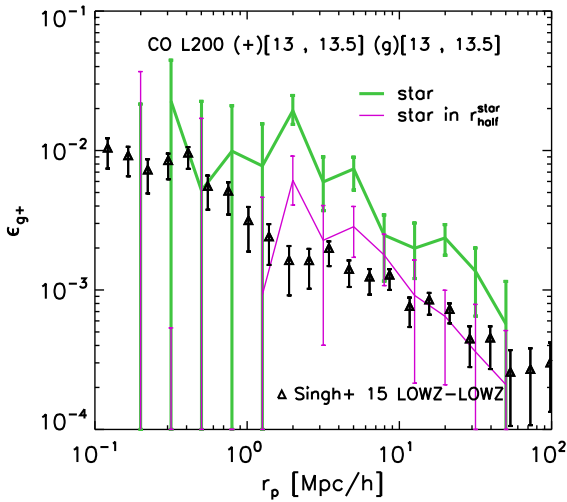} 
 \end{center}
\caption{Values of the observationally accessible proxy for orientation-direction alignment, $\epsilon_{\rm g+}$  (see Eq.~\ref{eq:epsilon}), as a function of the projected separation, $r_{\rm p}$, form simulations considering all stars bound to the subhalo (green curves) and only stars within $r_{\rm half}^{\rm star}$ (magenta curve). The data points are the observational results from \citet{Singh14} for the LOWZ sample of LRG galaxies (their Fig.~19, note that in their work $\epsilon_{\rm g+}$ is denoted $\left< \gamma \right>$ for its direct connection with the shear). We consider pairs of galaxies that have a separation along the projected axis smaller than $100\hMpc$. The error bars on the curves indicate one sigma bootstrap errors. When we consider all the stars, the predicted alignment is stronger than observed. However, when we only use stars in the part of the galaxy that might typically be observed, we find good agreement with the data.}
\label{fig:lowz_epsilon_g_plus}
\end{figure}

Recent observational studies of the orientation-direction alignment in galaxy groups and clusters reported signals consistent with zero alignment \citep{Chisari14, Sifon15}.
Specifically, \citet{Sifon15} used a sample of $\approx14,000$ spectroscopically confirmed galaxy members of 90 galaxy clusters with median mass of $\newl(M_{\rm 200}/ [\Msun]) = 14.8 $ and median redshift of $ z = 0.14$, selected as part of MENeaCS \citep[Multi-Epoch Nearby Cluster Survey;][]{Sand12} and CCCP \citep[Canadian Cluster Comparison Project;][]{Hoekstra12}. 
They constrained the average ellipticity, within the host virial radius, to be $\epsilon_{\rm g+}= (- 3.7 \pm 2.7) \times 10^{-3}$ or $\epsilon_{\rm g+}= (0.4\times \pm 3.1)\times 10^{-3}$ depending on the shape estimation method employed.
\citet{Chisari14} measured galaxy alignments in 3099  photometrically-selected galaxy groups in the redshift range between $z = 0.1$ and $z = 0.4$ of masses $\newl(M_{\rm 200}/ [\Msun]) = 13 $ in SDSS Stripe 82 and constrained the alignments to similar values as \citet{Sifon15}.

The left panel of Fig.~\ref{fig:groups_cluster_scales_epsilon_g_plus} shows the value of $\epsilon_{\rm g+}$ calculated for the simulations using all the stellar particles of subhaloes for host masses and satellite masses that are roughly comparable to the range of masses explored in \citet{Chisari14} and \citet{ Sifon15}. We only consider pairs separated by less than  $2.5\hMpc$ along the projection axis to confine the measurement to the typical extent of massive bound structures. Within the virial radii of groups or clusters the statistical uncertainties are large.
The average values of $\epsilon_{\rm g+}$ for distances smaller than the host virial radii are $\approx 2-4 \times 10^{-2}$ with errors  of $\approx 0.1-2 \times 10^{-2} $, indicating positive alignment. We repeat the same analysis using only stars within $r_{\rm half}^{\rm star}$ (see right panel of Fig.~\ref{fig:groups_cluster_scales_epsilon_g_plus}). In this case the average value of $\epsilon_{\rm g+}$ for distances that are smaller than the host virial radius is consistent with zero, in agreement with the observations of \citet{Chisari14} and \citet{Sifon15}. Using deeper observations, in order to probe the lower surface brightness parts of satellite galaxies, could represent a way to reveal the alignment that is seen in observations when all stars bounded to subhaloes are considered.

Recently, \citet{Singh14} measured the relative alignment of SDSS-III BOSS DR11 LOWZ Luminous Red Galaxies (LRGs) in the redshift range $0.16 < z < 0.36$ observed spectroscopically in the BOSS survey \citep[][]{BOSS13}. As opposed to the case of galaxy groups and clusters, these measurements are obtained by integrating along the line of sight between $\pm 100$ Mpc. Furthermore, \citet{Singh14} reported the average halo masses of those galaxies, as obtained from galaxy-galaxy lensing analysis. We perform the same measurements as in \citet{Singh14} on our simulations. Given the observed halo mass ($\newl (M_{180}^{\rm mean}[\hMsun])= 13.2$) and the line of sight integration limits, we employ the cosmo-OWLS L200 in this analysis.

Fig.~\ref{fig:lowz_epsilon_g_plus} shows the values of $\epsilon_{\rm g+}(r_{\rm p})$ from our simulation 
together with the measurements from \citet{Singh14}. Note that we have used a halo mass bin ($13 < \newl(M^{\rm crit}_{\rm sub}/ [\hMsun]) < 13.5$) half a magnitude wide to obtain statistically robust measurements ($N_{\rm haloes} = 1677$).
As for the case of satellite galaxies in clusters, the agreement with observational results depends strongly on the subset of stars used to compute the galaxy orientations. When one considers all stars bound to  subhaloes, the values obtained for $\epsilon_{\rm g+}(r_{\rm p})$ are systematically higher than the values in observations, whereas broad agreement is found when using only stars inside $r_{\rm half}^{\rm star}$.

As noted before, when only stars within $r_{\rm half}^{\rm star}$ are considered, simulated galaxies exhibit rounder shapes than observed. Therefore, the results presented here may underestimate the values of $\epsilon_{\rm g+}(r_{\rm p})$. Thus, when more observationally motivated algorithms would be employed to analyse the simulations, it is not guaranteed that 
the agreement found in Fig.~\ref{fig:lowz_epsilon_g_plus} would still hold.

\section{Orientation-orientation alignment}
\label{Sec:II}

\begin{figure} \begin{center} 
\includegraphics[width=1.0\columnwidth]{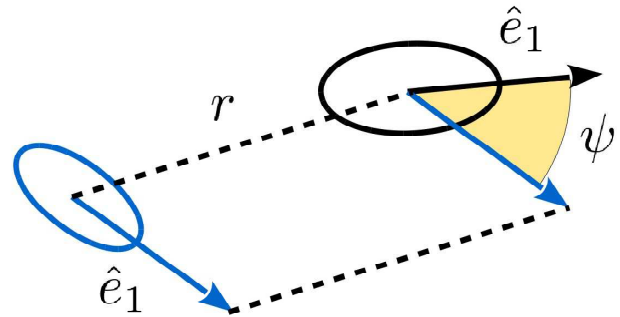} 
 \end{center}
\caption{Diagram of the angle $\psi$ formed between $\hat e_1$ of galaxy pairs at a distance $r$.
}
\label{fig:psi_diagram}
\end{figure}

In this section, we present the results for the II (intrinsic-intrinsic) term of the intrinsic alignment that is given by the angle between the orientations of different haloes.
We define $\psi$ as 
\begin{equation}
\psi(|\vec{r}|) = \arccos(|\hat e_1(\vec{x}) \cdot \hat e_1(\vec{x}+\vec{r})|).
\end{equation}
where $\hat e_1$ are the major eigenvectors of the 3D stellar distributions of a pair of galaxies separated by a 3D distance $r = |\vec{r}|$ (see Fig.~\ref{fig:psi_diagram}). 

\begin{figure} \begin{center} 
\includegraphics[width=1.0\columnwidth]{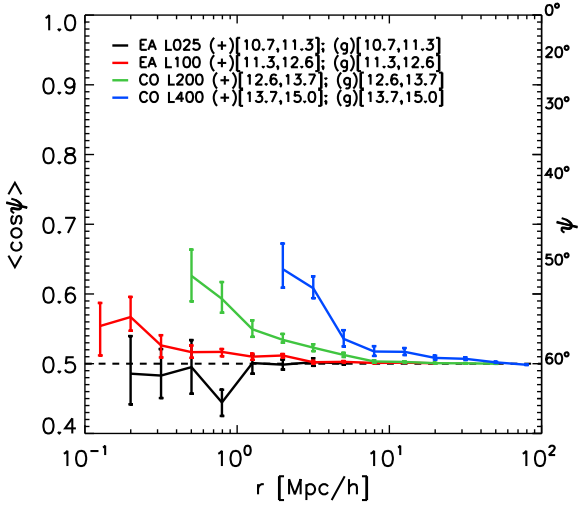} 
 \end{center}
\caption{Mean value of the cosine of the angle $\psi$  between the major axes of the stellar distributions of subhaloes as a function of their 3D separation. Each mass bin is taken from a different simulation. The minimum subhalo mass in every bin ensures that only haloes with more than 300 stellar particles are selected. Orientations are computed using all stars bound to the subhaloes. The orientation-orientation alignment decreases with distances and increases with mass. It is weaker than the orientation-direction alignment (cf. left panel of Fig.~\ref{fig:total_auto_mass_cos_phi}). 
}
\label{fig:cos_psi}
\end{figure}

Fig.~\ref{fig:cos_psi} shows the average value of the cosine of the angle $\psi$ for pairs of subhaloes with similar masses at a given 3D separation $r$ (in $\hMpc$).
Values are shown for four different choices of subhalo mass, where each mass bin is taken from a different simulation (see legend). 
To estimate the errors, we bootstrap the shape sample 100 times and take as 1-sigma error bars the 16th and the 84th percentile of the bootstrap distribution.
Values of $\left< \cos \psi \right>$ equal to 0.5 indicate a random distribution of galaxy orientations, whereas values of $\left< \cos \psi \right>$ higher than 0.5 indicate that on average galaxies are preferentially oriented in the same direction.

The alignment between the orientation of the stellar distribution decreases with distance and increases with subhalo mass.
Comparing with Fig.~\ref{fig:total_auto_mass_cos_phi}, the orientation-orientation alignment is systematically lower than the orientation-direction angle alignment. Beyond $50\hMpc$ the alignment is consistent with a random distribution, whereas in the orientation-direction case a positive alignment was found for scales up to $100 \hMpc$. This is suggestive of the \emph{direction} of nearby galaxies as being the main driver of the orientation-orientation alignment, as a weaker orientation-orientation alignment naturally stems from the dilution of the orientation-direction alignment. 

Similarly to $\epsilon_{\rm g+}$ (in Eq.~\ref{eq:epsilon}), we can define the projected orientation-orientation $\epsilon_{\rm ++}$ as:
\begin{equation}
\label{eq:epsilon_plus_plus}
\epsilon_{\rm ++}(r_p)=  \sum_{i\neq j\mid r_{p}}\frac{\epsilon^i_{+}\epsilon^j_{+}(j\mid i)}{N_{\rm pairs}},
\end{equation}
where $\epsilon_{+}$ is defined in Eq.~\ref{eq:epsilon_plus}. Galaxies are selected to have at least 300 star particles.

\begin{figure} \begin{center}
\includegraphics[width=1.0\columnwidth]{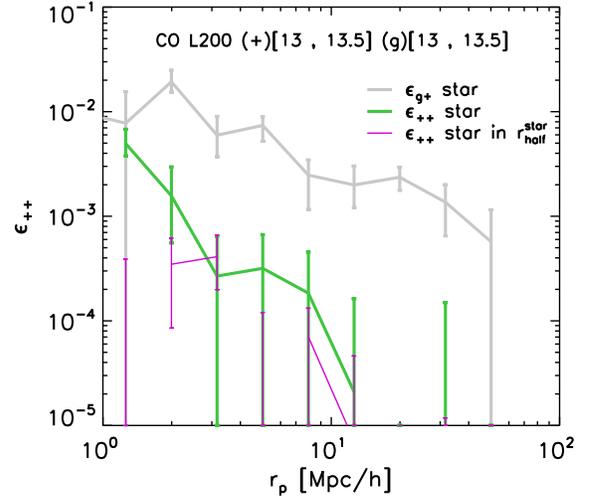} 
\end{center}
\caption{Dependence of $\epsilon_{++}$ (Eq.~\ref{eq:epsilon_plus_plus}), a measure of orientation-orientation alignment, obtained from the simulations using an integration limit of $100\hMpc$. Both the values for the whole stellar distribution (in green) and for the stars within $r_{\rm half}^{\rm star}$ are shown (in purple). The error bars indicates one sigma bootstrap errors. The results for $\epsilon_{\rm g+}$ (grey curve) are shown for comparison.}
\label{fig:lowz_lowz_epsilon_plus_plus}
\end{figure}

Fig.~\ref{fig:lowz_lowz_epsilon_plus_plus} shows the projected orientation-orientation alignment, $\epsilon_{++}$, for the same halo mass bin and integration limits as employed in Fig.~\ref{fig:lowz_epsilon_g_plus}. Green and magenta curves refer to the cases where one uses all stellar particles in subhaloes and only stellar particles confined within $r_{\rm half}^{\rm star}$, respectively. For comparison, $\epsilon_{\rm g+}( r_{\rm p})$ is overplotted in grey. As expected, the $\epsilon_{++}(r_{\rm p})$ profile has an overall lower normalization. Interestingly, $\epsilon_{++}(r_{\rm p})$ is steeper than $\epsilon_{g+}(r_{\rm p})$, although the significance of this trend is diminished by the noisy behaviour of the $\epsilon_{++}(r_{\rm p})$ profile.

The presence of a non-vanishing $\epsilon_{++}(r_{\rm p})$ profile reveals a net alignment of galaxies with the orientations of nearby galaxies, thus suggesting a potential II term in cosmic shear measurements for galaxies residing in haloes with masses $13 < \newl(M^{\rm crit}_{\rm sub}/ [\hMsun]) < 13.5$.

\section{Conclusions}
\label{Sec:Conclusions}
This paper reports the results of a systematic study of the orientation-direction and orientation-orientation alignment of galaxies in the EAGLE \citep{Schaye14,Crain15} and cosmo-OWLS \citep{LeBrun14,McCarthy14} hydro-cosmological simulations.
The combination of these state-of-the-art hydro-cosmological simulations enables us to span four orders of magnitude in subhalo mass ($10.7 \le \newl({M_{\rm sub}/ [\hMsun]}) \le 15$) and a wide range of galaxy separations ($-1 \le \newl(r/[\hMpc]) \le 2$).
For the orientation-direction alignment we define the galaxy orientation to be the major eigenvector of the inertia tensor of the distribution of stars in the subhalo, $\hat e_1$.
We then compute the mean values of the angle $\phi$ between $\hat e_1$ and the normalized separation vector, $\hat d$, towards a neighbouring galaxy at the distance $r$, for galaxies in different subhalo mass bins. In the case of orientation-orientation alignment, we compute the mean value of $\psi$, the angle between the major axes $\hat e_1$ of galaxy pairs separated by a distance $r$.

Our key findings are:
\begin{itemize}
\item Subhalo mass affects the strength of the orientation-direction alignment of galaxies for separations up to tens of $\Mpc$, but for distances greater than approximately ten times the subhalo radius the dependence on mass becomes insignificant. The strength of the signal is consistent with no orientation-direction alignment for separations $\gg 100$ times the subhalo radius (Figs.~\ref{fig:total_auto_mass_cos_phi}-\ref{fig:total_cross_mass_cos_phi}).

\item The difference between the orientation-direction alignment obtained using the dark matter, all the stars or the stars within $r^{\rm star}_{\rm half}$ to define galaxy orientations, could account for the common findings reported in the literature of galaxy alignment being systematically stronger in simulations than reported by observational studies (Fig.~\ref{fig:total_dm_star_halfstar_cos_phi}).  Since observations are limited to the shape and orientation of the region of a galaxy above a limit surface brightness, simulations have to employ proxies for the extent of this region.

\item At a fixed mass, subhaloes hosting more aspherical or prolate stellar distributions show stronger orientation-direction alignment (Fig.~\ref{fig:morpho_cos_phi}).

\item The distribution of satellites is significantly aligned with the orientation of the central galaxy for separations up to $100$ times the virial radius of the host halo ($r_{\rm 200}^{\rm crit}$), within $10 \, r_{\rm 200}^{\rm crit}$ higher-mass satellites show substantially stronger alignment (Fig.~\ref{fig:CenSat_cos_phi}).

\item Satellites are radially aligned towards the directions of the centrals. The strength of the alignment of satellites decreases with radius but is insensitive to the mass of the host halo (Fig.~\ref{fig:SatCen_cos_phi}).

\item Predictions for the radial profile of the projected orientation-direction alignment of galaxies, $\epsilon_{\rm g+}(r_{\rm p})$, depend on the subset of stars used to measure galaxy orientations. When only stars within $r_{\rm half}^{\rm star}$ are used, we find agreement between results from our simulations and recent observations from \citet{Sifon15} and \citet{Singh14}(see Figs.~\ref{fig:groups_cluster_scales_epsilon_g_plus} and \ref{fig:lowz_epsilon_g_plus}, respectively).

\item Predictions for the radial profile of the orientation-orientation alignment of galaxies, $\epsilon_{\rm ++}(r_{\rm p})$, are systematically lower than those for the orientation-direction alignment, $\epsilon_{\rm g+}(r_{\rm p})$, and have a steeper radial dependence (Figs.~\ref{fig:cos_psi} and \ref{fig:lowz_lowz_epsilon_plus_plus}). 
Although low, the non vanishing $\epsilon_{++}(r_{\rm p})$ profile reveals a net alignment of galaxies with the orientations of nearby galaxies, thus suggesting a potential intrinsic-intrinsic term in cosmic shear measurements for galaxies residing in haloes with masses $13 < \newl(M^{\rm crit}_{\rm sub}/ [\hMsun]) < 13.5$.

\end{itemize}

For a direct comparison with the observations, in order to validate the models or to explain the observations, particular care has to be taken to compare the same quantities in simulations and observations.
A future development of this work will be to extend the comparison with observations further by using the same selection criteria for luminosity, colour, and morphology in the simulations and in the observations.

The strength of galaxy alignments depends strongly on the subset of stars that are used to measure the orientations of galaxies and it is always weaker than the alignment of the dark matter components. Thus, alignment models that use halo orientation as a direct proxy for galaxy orientation will overestimate the impact of intrinsic galaxy alignments on weak lensing analyses.


\section*{Acknowledgements}  
\label{sec:acknowledgements}
We thank the anonymous referee for insightful comments that helped improve the manuscript. This work used the DiRAC Data Centric system at Durham University, operated by the Institute for Computational Cosmology on behalf of the STFC DiRAC HPC Facility (www.dirac.ac.uk). This equipment was funded by BIS National  E-infrastructure capital grant ST/K00042X/1, STFC capital grant ST/H008519/1, and STFC DiRAC Operations grant ST/K003267/1 and Durham University. DiRAC is part of the National E-Infrastructure. We also gratefully acknowledge PRACE for awarding us access to the resource Curie based in France at Tr\`es Grand Centre de Calcul. This work was sponsored by the Dutch National Computing Facilities
Foundation (NCF) for the use of supercomputer facilities, with financial support from the Netherlands Organization for Scientific
Research (NWO). The research was supported in part by the European Research Council under the European Union's Seventh Framework
Programme (FP7/2007-2013) / ERC Grant agreements 278594-GasAroundGalaxies, and 321334 dustygal. This research was supported by ERC FP7 grant 279396 and ERC FP7 grant 278594. RAC is a Royal Society University Research Fellow. TT acknowledge the Interuniversity Attraction Poles Programme initiated by the Belgian Science Policy Office ([AP P7/08 CHARM])


\bibliographystyle{mn2e}
\bibliography{paper} 
\expandafter\ifx\csname natexlab\endcsname\relax\def\natexlab#1{#1}\fi


\label{lastpage}
\end{document}